\documentclass[10pt,preprint]{aastex}
\usepackage{amsmath}
\pdfoutput=1

\shorttitle{Localizing coalescing black hole binaries}
\shortauthors{Lang \& Hughes}

\begin{document}

\title{Localizing coalescing massive black hole binaries with
gravitational waves}

\author{Ryan N.\ Lang and Scott A.\ Hughes}

\affil{Department of Physics and Kavli Institute for Astrophysics and Space Research, Massachusetts Institute of Technology, 77 Massachusetts Avenue, Cambridge, MA 02139}

\begin{abstract}
Massive black hole binary coalescences are prime targets for
space-based gravitational wave (GW) observatories such as {\it LISA}.  
GW measurements can localize the position of a
coalescing binary on the sky to an ellipse with a major axis of
a few tens of arcminutes to a few degrees, depending on 
source redshift, and a minor axis which is $2 - 4$
times smaller.  Neglecting weak gravitational lensing,
the GWs would also determine the source's luminosity distance to
better than percent accuracy for close sources, degrading to several
percent for more distant sources.  Weak lensing cannot, in fact, be
neglected and is expected to limit the accuracy with which distances
can be fixed to errors no less than a few percent.  Assuming a
well-measured cosmology, the source's redshift could be inferred with
similar accuracy.  GWs alone can thus pinpoint a binary to a
three-dimensional ``pixel'' which can help guide searches for the
hosts of these events.  We examine the time evolution of this pixel, studying it at merger and at several intervals before
merger.  One day before merger, the major axis of the error ellipse 
is typically larger
than its final value by a factor of $\sim 1.5-6$.  The
minor axis is larger by a factor of $\sim 2-9$, and, neglecting lensing, the error
in the luminosity distance is larger by a factor of $\sim 1.5-7$.  This
large change over a short period of time is due to spin-induced precession, 
which is strongest in the final days before
merger.  The evolution is slower as we go back further in time.  
For $z = 1$, we find that GWs will localize a
coalescing binary to within $\sim 10\ \mathrm{deg}^2$ as
early as a month prior to merger and determine distance (and hence
redshift) to several percent.
\end{abstract}


\section{Introduction}
\label{sec:intro}

Among the most important sources of gravitational waves (GWs) in the
low-frequency band of space-based detectors are the coalescences of
massive black hole binaries (MBHBs).  Binaries containing black holes
with masses in the range $10^4 - 10^7\,M_\odot$ are predicted to form
through the hierarchical growth of structure as dark matter halos (and
the galaxies they host) repeatedly merge; see, for example,
{\citet{svh07}} and {\citet{mhsa07}} for recent discussion.  The {\it Laser
Interferometer Space Antenna} ({\it LISA}) is being designed to have a
sensitivity that would allow detailed measurement of the waves from
these binaries.  ``Intrinsic'' parameters --- the masses and spins of
the black holes which compose the binary --- should be determined with
very high accuracy, with relative errors typically $\sim 10^{-3}$ to 
$10^{-1}$, depending on system mass and redshift; see Lang \& Hughes
(2006, hereafter Paper I) for recent discussion.  By measuring an
ensemble of coalescences over a range of redshifts, MBHB GWs may serve
as a kind of structure tracer, tracking the growth and spin
evolution of black holes over cosmic time.

``Extrinsic'' system parameters, describing a binary's location and
orientation relative to the detector, are also determined by measuring
its GWs.  In Paper I, we showed that a binary's position on the sky
can be localized at $z = 1$ to an ellipse with a major axis of a few
tens of arcminutes and a minor axis a factor of $2-4$ smaller.  At
higher redshift ($z = 3-5$), these values degrade by a factor of a
few, reaching a few degrees in the long direction and tens of
arcminutes to a degree or two in the short one.  We also found that,
neglecting errors due to weak gravitational lensing, a source's
luminosity distance typically can be determined to better $1\%$
at low redshift ($z = 1$), degrading to several percent at
higher redshift ($z = 3-5$).

The intrinsic ability of GWs to determine the distance to a coalescing
binary is phenomenal.  Coalescing MBHB systems constitute exquisitely
well calibrated distance measures, with the calibration provided by
general relativity.  Unfortunately, this percent-level or better
accuracy could only be achieved if we measured MBHB coalescences in an
empty universe.  In our universe, weak lensing will magnify or
demagnify the GWs, and we will infer a luminosity distance smaller
(for magnification) or larger (for demagnification) than the true
value.  This phenomenon affects all high-redshift standard candles.  Its
impact on Type Ia supernovae in particular has been discussed in
detail {\citep{frieman97,hw98,h98}}.  It will not be possible to
correct for this effect {\citep{dhcf03}} since much of the lensing
``noise'' arises from structure on subarcminute scales that is not
probed by shear maps (which map the distribution of matter on scales
greater than $1^\prime$ or so).  Since we will not know the extent
of the magnification when we measure MBHB waves, we must simply accept
the fact that lensing introduces a dispersion of several percent in determining
the distance to these GW events (see, e.g., Wang et al. [2002] to
compute this dispersion as a function of redshift).  When we quote
distance measurement errors, we will typically quote only the
intrinsic GW measurement error, neglecting lensing's impact.  When
the intrinsic GW distance error is $\lesssim 5\%$, lensing will blur it
 to the several percent level.

Note that a source's redshift $z$ {\it cannot} be directly determined
using only GWs.  Gravitational wave measurements infer system parameters 
through their
impact on certain dynamical timescales, such as orbital frequencies
and the rate at which these frequencies evolve.  Since these time
scales all suffer cosmological redshift, $z$ is degenerate with other
parameters.  For example, any mass parameter $m$ is actually measured
as $(1 + z)m$.  However, if the binary's luminosity distance is
determined, its redshift can then be inferred by assuming a
cosmography.  For most binaries, the redshift can be determined to
several percent (with an error budget dominated by gravitational
lensing\footnote{At redshifts $z \lesssim 0.3$, the error is actually
dominated by peculiar velocity effects {\citep{kfhm06}}; however, the
event rate is probably negligible at such low redshifts.  As such, we
will focus on gravitational lensing as the main source of systematic
redshift error.}).  We thus expect that GW measurements will locate a
binary to within a three-dimensional ``GW pixel'' which at $z = 1$ has a
cross-sectional area of $\sim 10^{-2}$ to $10^{-1}\ \mathrm{deg}^2$ and a
depth $\Delta z/z \sim $ several percent.

It is anticipated that there will be great interest in searching the
GW pixel for electromagnetic (e.g., optical, X-ray, radio)
counterparts to MBHB GW events.  Finding such a counterpart would be
much easier if galactic activity were catalyzed in association with the
coalescence {\citep{kfhm06}}.  The nature of that activity is likely
to depend rather strongly on the mass of the coalescing system
\citep{dssch06}.  For example, {\citet{an02}} predict strong outflows
and galactic activity prior to the final black hole merger as the
smaller member of the binary drives gas onto the larger member,
consistent with the high-mass ($M_{\rm tot} \gtrsim 10^7\,M_\odot$)
predictions of \citet{dssch06}.  {\citet{mp05}} describe an X-ray
afterglow that would ignite after gas refills the volume that is swept
clean by the coalescing binary; Dotti et al.\ predict this outcome for
smaller systems ($M_{\rm tot} \lesssim \mbox{several}\times
10^6\,M_\odot$).  Recent work by {\citet{bp07}} suggests that the
final burst of radiation from a coalescing binary (which can convert
$\sim 10\%$ of the system's mass to GWs very suddenly) may excite
radial waves, and consequently electromagnetic variability, in an accretion 
disk due to the quick change in the disk's Keplerian potential.
Such a signature may be essentially mass independent.  On the other
hand, the coalescence may be electromagnetically quiet, in which case
we face the potentially daunting task of searching the three-dimensional
pixel for galaxies with morphology consistent with a (relatively)
recent merger, or that have a central velocity dispersion $\sigma$
consistent with the inferred final black hole mass (assuming that the
$M_{\rm BH}-\sigma$ relation [Ferrarese \& Merritt 2000; Gebhardt et al. 2000] holds at the redshift
of these sources, and so soon after merger).

If the host galaxy or some other counterpart can be identified, we
could then contemplate combining GW information with electromagnetic
data.  For instance, combining {\it LISA} mass measurements with the
luminosity of the counterpart may allow us to directly measure the
Eddington ratio $L/L_{\mathrm{Edd}}$ {\citep{kfhm06}}.  Identifying a
counterpart would also allow us to more accurately characterize the
system.  For example, much of the intrinsic luminosity distance error
is due to correlations between distance and sky position.  Finding an
electromagnetic counterpart essentially determines a binary's position
precisely, breaking those correlations.  Previous studies have found
that intrinsic distance error can be reduced by almost an order of
magnitude if the position is known {\citep{h02,hh05}}.  (Lensing
errors still dominate in such a case, so that the distance remains
determined only at the few percent level.)  A counterpart may also
make it possible to simultaneously determine a source's luminosity
distance and redshift.  Such a ``standard siren'' (the GW analog of a
standard candle) would very usefully complement other high-redshift
standard candles {\citep{hh05}}, such as Type Ia supernovae
{\citep{p93, rpk95, wgap03}}.  A direct measurement of redshift will
also break the mass-redshift degeneracy more accurately than can be
done with just the luminosity distance and some assumed cosmological
parameters.  Breaking this degeneracy is critical when studying the
growth of black holes with cosmic time {\citep{h02}}.

Many analyses {\citep{c98,h02,v04,bbw05,hh05}} have quantified how
well {\it LISA} can determine MBHB parameters, including sky position
and distance, using maximum likelihood Fisher matrix estimation. Our
results from Paper I, given earlier, include the effects of
``spin-induced precession'' --- precession of both the orbital plane
and the individual spins of the black holes due to post-Newtonian
spin-interaction effects.  A significant result from that analysis is
that spin-induced precession improves sky position accuracy by about
half an order of magnitude in each direction versus previous analyses.
This result can be partially understood as due to the breaking of a
degeneracy between position and orientation angles: thanks to
precession, the binary's orientation evolves with time and can be
untangled from sky position.  This effect was already known due to
pioneering work by {\citet{v04}}; by taking the analysis to higher
order and considering a broader range of sources, we were able to show
that this improvement held for essentially all astrophysically
interesting MBHB sources.

The purpose of this paper is to examine the localization of MBHB
systems more thoroughly, in particular how the GW pixel evolves as the
final merger is approached.  Paper I only presented results for
measurements that proceed all the way to merger.  It will clearly be
of some interest to monitor potential hosts for the binary event some
time before the merger happens; if nothing else, telescopes will need
prior warning to schedule observing campaigns.  Understanding the rate
at which localization evolves can also have an important impact on the
design of the {\it LISA} mission, clarifying how often it will be
necessary to downlink data about MBHB systems in order to effectively
guide surveys.

Our main goal is to understand for what range of masses and redshifts prior
localization of a binary using GWs will be possible.  A previous
analysis by Kocsis et al.\ (2007b; hereafter K07) also examined this
problem in great detail, but without including the impact of
spin-induced precession.  One of our goals is to see to what extent
precession physics changes the conclusions of K07.  We find that
precession has a fairly small impact on the time evolution of the GW
pixel except in the last few days before the final merger, at which
point its impact can be tremendous.  Precession typically changes the
area of the sky position error ellipse by a factor of $\sim 3-10$ (up to
$\sim 60$ in extreme cases) in just the final day.  This is in accord
with the predictions of K07 (and even earlier predictions by N. Cornish
2005, unpublished).

The structure of this paper is as follows.  First, in \S\
{\ref{sec:background}}, we briefly review the basics of the MBHB
gravitational waveform and the parameter estimation formalism that we
use; this section is essentially a synopsis of relevant material from
Paper I.  Section {\ref{sec:intrinsicGW}} reviews the form of the GWs
that we use in our analysis, while \S\ {\ref{sec:extrinsicGW}}
describes how those waves are measured by the {\it LISA}
constellation.  We describe the measurement formalism we use in
\S\ {\ref{sec:formalism}}.  In \S\ {\ref{sec:review}}, we
summarize our results from Paper I regarding the final localization
accuracy that {\it LISA} can expect to achieve.

We turn to a detailed discussion of the time evolution of the GW pixel
in \S\ {\ref{sec:results}}.  We begin by summarizing the key ideas
behind the ``harmonic mode decomposition'' of K07 in \S\
{\ref{sec:khmf}}.  This technique cleverly allows calculation of the
GW pixel and its time evolution with much less computational effort
than our method (albeit without including the impact of spin
precession).  Unfortunately, we have discovered that some of the
approximations used by K07 introduce a systematic underestimate of
the final sky position error by a factor of $2 - 4$ or more in angle;
the approximations are much more reliable a week or more prior to the
black holes' final merger.  Modulo this underestimate, the K07
results agree well with a version of our code which does not include
spin precession (particularly a week or more in advance of merger,
when their underestimate is not severe).  K07 thus serves as a useful
point of comparison to establish the impact of precession on source
localization.

Section {\ref{sec:timeevolve}} is dedicated to our results, including
comparison to K07 when appropriate.  We find that all relevant
parameter errors decrease slowly with time until the last day before
merger, when they drop more dramatically.  This sudden drop is not
found in K07, nor is it present in a variant of our analysis that
ignores spin precession.  It clearly can be attributed to the impact
of precession on the waveform.  Before this last day, the major axis
is $\sim 1.5-6$ times, the minor axis $\sim 2-9$ times, 
and the intrinsic error in
the luminosity distance $D_L$ $\sim 1.5-7$ times bigger than at merger
for most binaries (i.e., all except the highest masses).  Going back
to one week (one month) before merger, these numbers change to $2-9$
($4-11$) for the major axis, $3-14$ ($5-24$) for the minor axis, and
$3-14$ ($5-18$) for the error in the luminosity distance.  As a
result, for $z = 1$, most binaries can be located within a few square
degrees a week before merger and $10\ \mathrm{deg}^2$ a month before
merger.  The intrinsic distance errors are also small enough this
early that $\Delta z/z$ remains dominated by gravitational lensing
errors of several percent.  Advanced localization of MBHB coalescences
thus seems plausible for these binaries; the situation is less
promising for sources at higher redshift.

As a corollary to our study of the time evolution, we also examine the
sky position dependence of errors (\S\ {\ref{sec:angdependence}}).
The errors depend strongly on the polar angle with respect to the
ecliptic, increasing in the ecliptic plane to as much as $35\%$ over
the median for the major axis, $85\%$ over the median for the minor
axis, and $15\%$ over the median for errors in the luminosity
distance.  The errors have a much weaker dependence on the azimuthal
angle.  When we convert to Galactic coordinates, we find that the best
localization regions appear to lie fairly far out of the Galactic
plane, offering hope that searches for counterparts will not be too badly
impacted by foreground contamination.

We conclude this paper in \S\ {\ref{sec:disc}}.  Besides summarizing
our results, we discuss shortcomings of this analysis and future work
which could help to better understand how well GWs can localize MBHB
sources.

Throughout the paper we set $G = c = 1$; a convenient conversion
factor in this system is $10^6 M_\odot = 4.92$ s.  When
discussing results, we always quote masses as they would be measured
in the rest frame of the source.  These masses must be multiplied by a
factor of $1 + z$ when used in any of the equations describing the
system's dynamics or its GWs (particularly the equations of \S\
{\ref{sec:background}}).  We convert between distance and redshift
using a spatially flat cosmology with $\Omega_\Lambda = 0.75$ (and
hence $\Omega_m = 0.25$) and Hubble constant $H_0 = 75\ {\rm
km}\ {\rm s}^{-1}\ {\rm Mpc}^{-1}$.

\section{Determining the sky position and distance accuracy of {\it LISA}}
\label{sec:background}

In this section, we present necessary background.  We begin by summarizing the
inspiral GW signal and its measurement by {\it LISA}.  We then
describe the parameter estimation formalism that we use in our code.
More details on these topics can be found in Paper I and references
therein.  Finally, we present a synopsis of the results from Paper I,
summarizing the {\it final} sky position and distance accuracy that
{\it LISA} can expect to achieve.  Detailed discussion of how this
accuracy evolves as a function of time is deferred to \S\
{\ref{sec:results}}.

\subsection{MBHB gravitational waves}
\label{sec:intrinsicGW}

As is by now traditional, we break the GWs generated by a coalescing
binary black hole system into three more or less distinct epochs: (1)
a slowly evolving {\it inspiral}, in which the black holes gradually
spiral toward each other as the orbit decays due to GW emission; (2)
a loud {\it merger}, in which the black holes come together and
form a single body; and (3) a {\it ringdown}, in which the merged
remnant of the binary settles down to its final state.  Our analysis
focuses on the inspiral, the most long-lived epoch of coalescence and
the epoch in which spin precession plays a major dynamical role.  The
inspiral's duration means that {\it LISA}'s position and orientation
changes considerably during the measurement; this, plus the impact of
precession, is largely responsible for pinning down the sky position
of coalescing sources.

The inspiral GW signal is modeled using the post-Newtonian
(PN) formalism {\citep{b06,bdiww95,ww96}}.  More
specifically, we use the ``restricted'' PN approximation.
The waveform can be written schematically as \citep{cf94}
\begin{equation}
h(t) = \mathrm{Re}\left[\sum_{x,m}
h^x_m(t)e^{im\Phi_{\mathrm{orb}}(t)}\right] \, ,
\label{eq:restrictedPN}
\end{equation}
where $x$ labels PN order, $m$ is a harmonic index, and
$\Phi_{\mathrm{orb}}(t)$ is the orbital phase.  In the restricted
PN case, we discard all amplitudes except the strongest
Newtonian quadrupole, $h_2^0$.  The phase, however, is computed to a
desired PN order (here the second one, or 2PN).  It has been recognized for
some time that important additional information is carried by
subleading harmonics (e.g., Hellings \& Moore 2003).  Recent work
{\citep{arunetal,ts08}} has confirmed that sky position and distance
are more sharply constrained when these terms are included in the wave
model (although there still appears to be some disagreement among
different groups as to the extent of the improvement).  We plan to
include these terms in future work, examining how higher harmonics and
precession combine to localize MBHB systems.

Spin-induced precession arises from geodetic and gravitomagnetic
influences on the angular momentum vectors present in the binary.
Averaged over an orbit, the leading-order spin precession equations
are {\citep{acst94,k95}}
\begin{eqnarray}
\mathbf{\dot{S}}_1 &=&
\frac{1}{r^3}\left[\left(2+\frac{3}{2}\frac{m_2}{m_1}\right)\mu
\sqrt{Mr} \mathbf{\hat{L}}\right] \times \mathbf{S}_1 +
\frac{1}{r^3}\left[\frac{1}{2}\mathbf{S}_2-\frac{3}{2}(\mathbf{S}_2\cdot
\mathbf{\hat{L}})\mathbf{\hat{L}}\right] \times \mathbf{S}_1 \, ,
\label{eq:S1dot}\\
\mathbf{\dot{S}}_2 &=&
\frac{1}{r^3}\left[\left(2+\frac{3}{2}\frac{m_1}{m_2}\right)\mu
\sqrt{Mr} \mathbf{\hat{L}}\right] \times \mathbf{S}_2 +
\frac{1}{r^3}\left[\frac{1}{2}\mathbf{S}_1-\frac{3}{2}(\mathbf{S}_1\cdot
\mathbf{\hat{L}})\mathbf{\hat{L}}\right] \times \mathbf{S}_2 \, ,
\label{eq:S2dot}
\end{eqnarray}
where $m_1$ and $m_2$ are the masses of holes 1 and 2, $M = m_1 + m_2$ 
is the total mass, $\mu = m_1m_2/M$ is the reduced mass,  
$\mathbf{\hat{L}}$ is the normal to the orbital plane, 
$\mathbf{S}_1$ and $\mathbf{S}_2$ are the spins of the holes,
and $r$ is their orbital separation in harmonic coordinates. 
Since total angular momentum is conserved
on short timescales (ignoring the inspiral), the orbital angular
momentum precesses to compensate.  These precessions drive several
modifications to the orbital phase.  One correction is due to
spin-orbit and spin-spin terms in the post-Newtonian expression for
orbital frequency: Precession makes these terms time dependent, adding
a new modulation to the GW phase.  Another correction arises from the
changing orientation of the orbital plane \citep{acst94}.

The waveform is described as two polarizations propagating in the
$-\mathbf{\hat{n}}$ direction, where $\mathbf{\hat{n}}$ gives the
direction to the binary on the sky:
\begin{eqnarray}
h_+(t) &=& \frac{2{\mathcal M}^{5/3}(\pi
f)^{2/3}}{D_L}[1+(\mathbf{\hat{L}}\cdot \mathbf{\hat{n}})^2]
\cos[\Phi(t) + \delta_p \Phi(t)] \, ,
\label{eq:hplus}\\
h_{\times}(t) &=& -\frac{4{\mathcal M}^{5/3}(\pi
f)^{2/3}}{D_L}(\mathbf{\hat{L}}\cdot \mathbf{\hat{n}}) \sin[\Phi(t) +
\delta_p \Phi(t)] \, ,
\label{eq:hcross}
\end{eqnarray}
where $f$ is the frequency of the waves, $D_L$ is the luminosity
distance,
$\Phi(t) = \int^t 2\pi f(t') dt'$ is the wave phase (twice the orbital
phase in the quadrupole approximation), and $\delta_p \Phi(t)$ is the
precessional correction to the orbital phase due to the changing
orientation of the orbital plane.  The ``chirp mass'' $\mathcal{M}$,
which controls the frequency evolution, or chirp, of the signal, is
given by $\mathcal{M} = \mu^{3/5}M^{2/5}$.  Note that the
relative weight of the polarizations is determined by the binary's
inclination to the line of sight $\iota$, where $\cos \iota =
\mathbf{\hat{L}}\cdot \mathbf{\hat{n}}$.

\subsection{MBHB gravitational waves as measured by {\it LISA}}
\label{sec:extrinsicGW}

The baseline {\it LISA} mission consists of three spacecraft arranged
in an equilateral triangle with arms of length $L = 5 \times 10^6$ km.
The center of this constellation orbits the Sun $20^\circ$ behind the
earth, with the triangle oriented at $60^\circ$ to the ecliptic.  The
individual spacecraft thus orbit in different planes, causing the
triangle to ``roll'' about itself in its orbit.  To model {\it LISA}'s
response to a signal, we use the long wavelength ($\lambda \gg L$)
approximation introduced by \citet{c98}.  The three arms of the
triangle act as a pair of two-arm detectors; the signals from the
three arms are combined in such a way that the noise spectrum in each
``synthesized'' detector is independent of the other.  Each detector
measures both GW polarizations, weighted by antenna response functions
$F_i^+(t)$ and $F_i^\times (t)$, where $i \in \{{\rm I, II}\}$ labels the
synthesized detector.  These functions depend on the source's sky
position as measured in the {\it detector} frame.  Because of the
constellation's rolling orbital motion, these functions are time
dependent.  The motion around the Sun also causes a Doppler shift in
the signal's frequency.  This time dependence allows the sky position
of the source to be measured.

An additional time dependence is produced by the precession of the
binary's orbital plane.  We saw above that the polarizations strongly
depend on the binary's inclination to the line of sight $\iota$ (via
$\cos\iota = \mathbf{\hat{L}}\cdot\mathbf{\hat{n}}$).  The antenna
pattern functions $F_i^{+,\times}(t)$ also depend on the component of
$\mathbf{\hat{L}}$ that is perpendicular to the line of sight.  Since
precession changes the direction of the orbital plane, the measured
polarizations would vary in time even if the detector were static.
The timescale of these precessions is much shorter than the {\it
LISA} orbital timescale of 1 yr, so they provide a great deal of
additional information about sky position.

We perform the parameter estimation analysis in the frequency domain, so
we Fourier transform the signal using the stationary phase approximation.  The final measured waveform is given by
\begin{equation}
\tilde{h}_i(f) = \sqrt{\frac{5}{96}}\frac{\pi^{-2/3} {\mathcal
M}^{5/6}}{D_L}A_{\mathrm{pol}, i}[t(f)] f^{-7/6}
e^{i(\Psi(f) - \varphi_{\mathrm{pol},i}[t(f)] -
\varphi_D[t(f)] - \delta_p\Phi[t(f)])} \, ,
\label{eq:freqdomainsignal}
\end{equation}
where $A_{\mathrm{pol}, i}(t)$ is the ``polarization amplitude,''
$\varphi_{\mathrm{pol},i}(t)$ is the ``polarization phase,'' and
$\varphi_D(t)$ is the Doppler phase.  The first two encode modulations
due to the relative polarization weightings and antenna pattern functions; the
latter is due to the Doppler shift in the GW frequency.  Full
expressions for these functions can be found in Paper I.  To second
post-Newtonian order, the phase $\Psi(f)$ is given by {\citep{pw95}}
\begin{eqnarray}
\Psi(f) &=& 2\pi f t_c - \Phi_c - \frac{\pi}{4} + \frac{3}{128}(\pi
{\mathcal M} f)^{-5/3} \left[1 +
\frac{20}{9}\left(\frac{743}{336} + \frac{11}{4}\eta \right)(\pi
Mf)^{2/3}\right.
\nonumber\\
& &\left. - 4(4\pi - \beta)(\pi Mf) + 10\left(\frac{3058673}{1016064}
+ \frac{5429}{1008}\eta + \frac{617}{144}\eta^2 - \sigma \right)(\pi
Mf)^{4/3}\right] \, ,
\label{eq:PNpsi}
\end{eqnarray}
where $\eta = \mu/M$, $t_c$ is the ``merger time'' (the time at which
$f$ formally diverges in the post-Newtonian framework), $\Phi_c$ is
the phase at $t_c$, and $\beta$ and $\sigma$ are ``spin-orbit'' and
``spin-spin'' parameters, respectively, encoding the manner in which
interactions between the spins and the orbits impact the phase.  See
Paper I for precise definitions.  For our purposes, the most important
property of $\beta$ and $\sigma$ is that they oscillate due to
precession, modulating the phase $\Psi(f)$.  Note that waveform
(\ref{eq:freqdomainsignal}) depends on 15 parameters: 2 masses, 6
initial spin components, 1 distance, 2 sky angles, 2 initial orbital
plane angles, $t_c$, and $\Phi_c$.

The end of inspiral/beginning of merger is a somewhat ad hoc and
fuzzy boundary.  Indeed, recent numerical computations have shown that
the GWs produced by a binary that coalesces into a single body do not show
any particular special feature as the black holes come together,
instead smoothly chirping through this transition
\citep{bcp07,panetal08}.  Since we are not including numerical merger
waves in our analysis, we require some point to terminate our
post-Newtonian expansion.  Most studies show that the inspiral comes
to an end when the separation of the bodies in harmonic coordinates is
roughly $r \sim 6M$; at this point, the system's GW frequency is
approximately given by
\begin{equation}
f_{\mathrm{merge}} \simeq \frac{2}{2\pi}\Omega_{\mathrm{Kepler}}(r =
6M) \simeq \frac{0.02}{M} \, ,
\label{eq:fmerge}
\end{equation}
where $\Omega_{\mathrm{Kepler}} = (M/r^3)^{1/2}$ is the Keplerian orbital 
angular frequency.
(The factor $1/2\pi$ converts from angular frequency to frequency; the
additional factor of $2$ accounts for the quadrupolar nature of
gravitational waves.)  We use equation (\ref{eq:fmerge}) throughout our
analysis to terminate the inspiral.

\subsection{Summary of parameter measurement formalism}
\label{sec:formalism}

To estimate how well {\it LISA} will be able to measure MBHB
parameters, we use the maximum likelihood formalism developed by \citet{f92}.  The GW signal $h_i(t)$ described above is corrupted by noise
to produce a measured signal $s_i(t) = h_i(t) + n_i(t)$.  We assume
the noise $n_i(t)$ is zero mean, wide-sense stationary, Gaussian, and
uncorrelated between the two detectors ($i \in \{{\rm I, II}\}$).  It can
be characterized by its (one sided) power spectral density (PSD) $S_n(f)$, which is described in detail in \S\ IIIB of Paper I.  The PSD
includes instrumental noise intrinsic to the detector, based on the
model described by \citet{lhh00} and implemented on the
World Wide Web\footnote{See http://www.srl.caltech.edu/\~{}shane/sensitivity/.}, with
important numerical factors described in \S\ IIC of \citet{bbw05}.
It also includes astrophysical noise arising from a confused
background of Galactic \citep{nyz01} and extragalactic \citep{fp03}
white dwarf binaries.  With the noise PSD, we can define the inner product between two signals $a$ and $b$ as 
\begin{equation}
(a|b) = 4 \, \mathrm{Re} \left[\int_0^{\infty} df
\frac{\tilde{a}^*(f)\tilde{b}(f)}{S_n(f)}\right]
= 2 \int_0^{\infty} df \frac{\tilde{a}^*(f)\tilde{b}(f) +
\tilde{a}(f)\tilde{b}^*(f)}{S_n(f)}  \, .
\label{eq:innerproduct}
\end{equation}

We write a GW as $h(\boldsymbol{\theta})$, where the components of the vector $\boldsymbol{\theta}$ represent the 15 parameters describing MBHB waves.  We assume that a detection has occurred --- that a signal with particular parameters $\boldsymbol{\tilde{\theta}}$ is present in the data --- and that maximum likelihood (ML) estimates $\boldsymbol{\hat{\theta}}$ \citep{cf94} of these parameters have been found.  The signal-to-noise ratio (S/N) is then given by
\begin{equation}
\mathrm{S/N} \approx (h(\boldsymbol{\hat{\theta}}) |
h(\boldsymbol{\hat{\theta}}))^{1/2} \, .
\label{eq:finalSNR}
\end{equation}

Our ultimate goal is to understand the errors in the ML parameters.  To do so, we examine the probability that the GW parameters are given by $\boldsymbol{\tilde{\theta}}$, expanded around the ML estimate $\boldsymbol{\hat{\theta}}$ \citep{cf94, pw95}:
\begin{equation}
p(\boldsymbol{\tilde{\theta}} | s) \propto \exp \left(-\frac{1}{2}\Gamma_{ab}\delta
\theta^a \delta \theta^b \right) \, ,
\label{eq:linearizedprob}
\end{equation}
where $\delta \theta^a = \tilde{\theta}^a - \hat{\theta}^a$ and
\begin{equation}
\Gamma_{ab} = \left(\frac{\partial h}{\partial \theta^a}\left|
\frac{\partial h}{\partial \theta^b}\right.\right) \, ,
\label{eq:fishermatrix}
\end{equation}
evaluated at $\boldsymbol{\theta} = \boldsymbol{\hat{\theta}}$, is
known as the Fisher matrix.  This expression holds for
large values of the S/N and is known as the Gaussian
approximation.  For the two-detector case, we can exploit the
independence of the noises to write the total Fisher matrix $\Gamma_{ab}^{\rm tot}$ as the sum of the individual Fisher matrices.  If we define the covariance
matrix $\Sigma^{ab} = (\Gamma_{\rm tot}^{-1})^{ab}$, we can then estimate the measurement error in some parameter
$\theta^a$,
\begin{equation}
\Delta \theta^a \equiv \sqrt{\langle(\delta \theta^a)^2\rangle} =
\sqrt{\Sigma^{aa}} \, ,
\label{eq:error}
\end{equation}
as well as the correlation between two parameters,
\begin{equation}
c^{ab} \equiv \frac{\langle\delta \theta^a \delta
\theta^b\rangle}{\Delta \theta^a \Delta \theta^b} =
\frac{\Sigma^{ab}}{\sqrt{\Sigma^{aa}\Sigma^{bb}}} \, .
\label{eq:correlation}
\end{equation}

We are particularly interested in the sky position, which we parameterize by the coordinates
($\bar{\mu}_N = \cos \bar{\theta}_N$, $\bar{\phi}_N$).\footnote{The bar
over the angles conforms to the notation in Paper I for quantities
measured in the solar system barycenter frame, as opposed to those
measured in the time-varying detector frame.}  We want to convert
from errors in these two parameters to an error ellipse on the sky.
To do so, we first perform a change of coordinates from $\bar{\mu}_N$
to $\bar{\theta}_N$.  For small deviations from the ML estimate, we
have $\delta\bar{\theta}_N = (d\bar{\theta}_N/d\bar{\mu}_N)\delta
\bar{\mu}_N = -\delta \bar{\mu}_N/\sin \bar{\theta}_N$.  Next, we
recognize that due to the geometric properties of the sphere, the
same $\delta \bar{\phi}_N$ corresponds to a different ``proper'' angle
depending on the value of $\bar{\theta}_N$: $\delta \bar{\phi}_N^p =
\sin \bar{\theta}_N \delta \bar{\phi}_N$.  With these modifications,
equation (\ref{eq:linearizedprob}) becomes
\begin{equation}
p(\boldsymbol{\tilde{\theta}}|s) \propto \exp
\left(-\frac{1}{2}\Gamma^p_{a^\prime b^\prime}\delta
\theta^{a^\prime}_p \delta \theta^{b^\prime}_p \right) \, .
\label{eq:properundiagonalized}
\end{equation}
Here $\delta \theta^{a^\prime}_p \equiv
(\delta\bar\theta_N,\delta\bar\phi_N^p)$ denotes the proper errors
accounting for the metric of the sphere, and $\Gamma^p_{a^\prime
b^\prime}$ represents the equivalent Fisher matrix with all conversion
factors absorbed inside:
\begin{align}
\Gamma^p_{\bar{\theta}_N \bar{\theta}_N} &= \sin^2 \bar{\theta}_N
\Gamma_{\bar{\mu}_N \bar{\mu}_N} \, , \\
\Gamma^p_{\bar{\phi}_N \bar{\phi}_N} &= \csc^2 \bar{\theta}_N
\Gamma_{\bar{\phi}_N \bar{\phi}_N} \, , \\
\Gamma^p_{\bar{\theta}_N \bar{\phi}_N} &= \Gamma^p_{\bar{\phi}_N
\bar{\theta}_N} = (-\sin \bar{\theta}_N)(\csc \bar{\theta}_N)\Gamma_{\bar{\mu}_N \bar{\phi}_N} = -\Gamma_{\bar{\mu}_N \bar{\phi}_N} \, , 
\label{eq:properfisher}
\end{align}
and so on for the rest of the elements.
The inverse of this matrix is the proper covariance matrix,
$\Sigma_p^{a^\prime b^\prime}$.  Consider now just the $2 \times 2$ subspace
of the covariance matrix containing the sky position variables.
Let the eigenvalues of this subspace be $\lambda_\pm$.
If we define the error ellipse such that the probability that the
source lies outside of it is $e^{-1}$ (corresponding to a $\approx
63\%$ confidence interval), then the major and minor axes are given by
$2a = 2(2\lambda_+)^{1/2}$ and $2b = 2(2\lambda_-)^{1/2}$, respectively.  
Expressed in terms of the original covariance matrix, these are
\begin{equation}
2\left[ \vphantom{\sqrt{(\Sigma^{\bar{\mu}_N\bar{\mu}_N} -
\Sigma^{\bar{\phi}_N\bar{\phi}_N})^2 +
4(\Sigma^{\bar{\mu}_N\bar{\phi}_N})^2}}
\csc^2 \bar{\theta}_N \Sigma^{\bar{\mu}_N\bar{\mu}_N} + 
\sin^2 \bar{\theta}_N \Sigma^{\bar{\phi}_N\bar{\phi}_N}
\pm \sqrt{(\csc^2 \bar{\theta}_N \Sigma^{\bar{\mu}_N\bar{\mu}_N} -
\sin^2 \bar{\theta}_N \Sigma^{\bar{\phi}_N\bar{\phi}_N})^2 +
4(\Sigma^{\bar{\mu}_N\bar{\phi}_N})^2} \right]^{1/2} \, .
\label{eq:2a2b}
\end{equation}
Many previous analyses have reported the ellipse's area $\Delta
\Omega_N = \pi a b$ or $(\Delta\Omega_N)^{1/2}$, the side of a square of
equivalent area, as the sky position error {\citep{c98,v04,bbw05,hh05}}.
Information about the ellipse's shape, crucial input to coordinating
GW observations with telescopes, is not included in such a measure.
By examining both $2a$ and $2b$, this information is restored.

In an empty universe, the third dimension of our pixel would be
determined from the luminosity distance error, $\Delta D_L/D_L =
(\Sigma^{\ln D_L, \ln D_L})^{1/2}$.  As already discussed, this
quantity is often so small that weak gravitational lensing is
expected to dominate the distance error budget.  We thus expect
\begin{equation}
\frac{\Delta D_L}{D_L} \simeq \sqrt{\Sigma^{\ln D_L, \ln D_L} +
\sigma^2_{\rm lens}(z)} \simeq {\rm max}\left[\sqrt{\Sigma^{\ln D_L,
\ln D_L}}, \sigma_{\rm lens}(z)\right] \, ,
\end{equation}
where the lensing dispersion $\sigma_{\rm lens}(z)$ is described in,
for example, \citet{whm02}.  If cosmological parameter errors can be
neglected, then one typically finds that the redshift error is about
equal to the distance error {\citep{h02}}: $\Delta z/z \approx \Delta
D_L/D_L$, independent of redshift.

\subsection{Final position and distance accuracy}
\label{sec:review}

We conclude this section by reviewing the final accuracy with which
MBHB sky position and distance can be determined with GWs.  In Paper
I, we described the code which implements the measurement formalism
described in \S \S\ {\ref{sec:intrinsicGW}} -- {\ref{sec:formalism}}
above.  This formalism is coupled to a Monte Carlo engine: We specify
the rest frame masses and redshift and then randomly choose the sky
position, initial angular momentum and spin directions, and spin
magnitudes for each binary.  We also uniformly distribute the time
parameter $t_c$ of each binary over the assumed duration of the {\it
LISA} mission (which we take to be 3 yr).  Some binaries are
therefore already ``on'' when the mission begins, and as such may
radiate in band for less time than other binaries which enter the band
later.  The random distribution of merger times also means that the
relative azimuth between a binary's sky position and {\it LISA}'s
orbital position at merger, $\delta\phi = \bar{\phi}_N - \phi_{\it
LISA}(t_c)$, is itself randomly distributed, even if we choose to
focus on a particular $\bar{\phi}_N$.  We discuss the implications of
this in more detail in \S\ {\ref{sec:angdependence}}.  Computing the
errors for a large number of binaries ($10^4$ for each mass and
redshift set, unless otherwise stated) allows us to make statements
about the distribution of parameter errors for different mass and
redshift combinations.

Table {\ref{tab:finalvals}} summarizes the final accuracy of MBHB sky
position and distance measurements.  Here we show the {\it median}
measurement accuracy from many Monte Carlo surveys.  Several trends
are particularly noteworthy.  At all redshifts, the accuracy is worst
for the largest masses.  This is because the most massive systems are
in band for the least amount of time: The frequency at which inspiral
ends is inversely proportional to mass (see eq. [\ref{eq:fmerge}]),
and more massive systems move more rapidly from low frequency to high
frequency.  The short time these systems spend in band means that {\it
LISA} measures a relatively small number of modulations (whether
induced by the constellation's orbit or by spin precession).  Second,
note that the results for $m_1 = m_2$ tend to be less accurate than
results with similar total mass but for which $m_1 > m_2$.  The cause
of this phenomenon lies in precession equations (\ref{eq:S1dot})
and (\ref{eq:S2dot}): When $m_1 \ne m_2$, the two spins precess at
different rates, imposing richer modulations on the measured GWs.
Since $m_1 = m_2$ is a rather artificial limit, we expect that the
more accurate results for nonunity mass ratio will be the rule.

\begin{table}[p]
\begin{center}
\caption{Sky position and distance accuracy at merger}
\begin{tabular}{ccccccc}
\\
\tableline \tableline
$z$ &
$m_1\ (M_\odot)$ &
$m_2\ (M_\odot)$ &
$2a\ (\mathrm{arcmin})$ &
$2b\ (\mathrm{arcmin})$ &
$\Delta\Omega_N (\mathrm{deg}^2)$ &
$\Delta D_L/D_L$ \\

\tableline

1 & $10^5$ & $10^5$ & 27.3 & 13.3 & 0.0729 & 0.00398 \\ 
& $3\times 10^5$ & $10^5$ & 16.9 & 7.33 & 0.0233 & 0.00240 \\ 
& & $3\times 10^5$ & 23.3 & 11.8 & 0.0556 & 0.00357 \\ 
& $10^6$ & $10^5$ & 27.2 & 6.62 & 0.0235 & 0.00320 \\ 
& & $3\times 10^5$ & 31.3 & 13.2 & 0.0705 & 0.00393 \\ 
& & $10^6$ & 40.2 & 21.9 & 0.176 & 0.00560 \\
& $3\times 10^6$ & $3\times 10^5$ & 34.1 & 9.20 & 0.0445 & 0.00376 \\ 
& & $10^6$ & 32.3 & 14.7 & 0.0839 & 0.00419 \\ 
& & $3\times 10^6$ & 43.3 & 22.3 & 0.193 & 0.00689 \\ 
& $10^7$ & $10^6$ & 37.6 & 12.2 & 0.0670 & 0.00457 \\ 
& & $3\times 10^6$ & 42.1 & 19.0 & 0.142 & 0.00610 \\ 
& & $10^7$ & 81.3 & 38.6 & 0.680 & 0.0250 \\ 

\tableline

3 & $10^5$ & $10^5$ & 81.0 & 40.8 & 0.665 & 0.0123 \\ 
& $3\times 10^5$ & $10^5$ & 92.5 & 39.5 & 0.656 & 0.0126 \\ 
& & $3\times 10^5$ & 142 & 75.7 & 2.15 & 0.0201 \\ 
& $10^6$ & $10^5$ & 141 & 36.6 & 0.739 & 0.0155 \\ 
& & $3\times 10^5$ & 129 & 56.7 & 1.25 & 0.0161 \\ 
& & $10^6$ & 158 & 84.3 & 2.64 & 0.0237 \\ 
& $3\times 10^6$ & $3\times 10^5$ & 132 & 40.3 & 0.751 & 0.0153 \\ 
& & $10^6$ & 142 & 64.6 & 1.65 & 0.0193 \\ 
& & $3\times 10^6$ & 224 & 111 & 5.08 & 0.0422 \\ 
& $10^7$ & $10^6$ & 206 & 78.5 & 2.74 & 0.0293 \\ 
& & $3\times 10^6$ & 297 & 152 & 9.40 & 0.0805 \\ 
& & $10^7$ & 2000 & 583 & 256 & 2.41 \\ 

\tableline

5 & $10^5$ & $10^5$ & 169 & 85.7 & 2.93 & 0.0260 \\ 
& $3\times 10^5$ & $10^5$ & 217 & 95.8 & 3.73 & 0.0284 \\ 
& & $3\times 10^5$ & 295 & 161 & 9.29 & 0.0409 \\ 
& $10^6$ & $10^5$ & 248 & 66.8 & 2.35 & 0.0273 \\ 
& & $3\times 10^5$ & 233 & 101 & 3.96 & 0.0294 \\ 
& & $10^6$ & 315 & 162 & 10.2 & 0.0501 \\ 
& $3\times 10^6$ & $3\times 10^5$ & 265 & 86.4 & 3.27 & 0.0318 \\ 
& & $10^6$ & 304 & 139 & 7.52 & 0.0436 \\ 
& & $3\times 10^6$ & 538 & 260 & 29.5 & 0.140 \\ 
& $10^7$ & $10^6$ & 577 & 290 & 31.9 & 0.124 \\ 
& & $3\times 10^6$ & 1720 & 621 & 234 & 1.24 \\ 
& & $10^7$ & 180000 & 29600 & $1.15\times10^6$ & 377 \\

\tableline
\end{tabular}

\tablecomments{Median errors in sky position and distance for binaries
at various masses and redshifts $z = 1$, $3$, and $5$; $2a$ is the
major axis of the sky position error ellipse, $2b$ is the minor axis, and $\Delta \Omega_N$ is the ellipse's area.  (Note that since all of these quantities are medians, it is {\it not} true that $\Delta \Omega_N = \pi a b$.)
At each mass and redshift point, $10^4$ binaries are drawn with
randomly distributed sky positions, initial orbit and spin orientations, spin magnitudes, and $t_c$.}
\label{tab:finalvals}
\end{center}
\end{table}

Independent of these trends, an important result is that 
MBHB systems are pinned down on
the sky fairly accurately at $z = 1$.  Modulo the higher mass
binaries, the median major axis of the sky position error ellipse is
typically about $15^\prime - 45^\prime$, and the median minor axis is
about $5^\prime - 20^\prime$, with a total ellipse area considerably
smaller than $1\ \mathrm{deg}^2$ (ranging from about 0.02 to 0.2 
$\mathrm{deg}^2$).  Sources at $z = 1$ are located accurately enough that one
can comfortably contemplate searching the GW error ellipse for MBHB
counterparts with future survey instruments, such as 
the Large Synoptic Survey Telescope (LSST; Tyson et al. 2002).

At higher redshift, positional accuracy degrades.  This is due to the
weakening of the signal with distance and to the redshifting of the
waves' spectrum, so that the signal tends to spend less time in band.
At $z = 3$, the major axis of the error ellipse is $\sim 1^\circ-4^\circ$
across, and the minor axis is $\sim 40^\prime-110^\prime$.  The total
area of this ellipse is $\sim 0.5 - 5\ \mathrm{deg}^2$.  At $z = 5$,
this degrades further to $\sim 3^\circ - 5^\circ$ for the major axis,
$\sim 1^\circ - 3^\circ$ for the minor axis, and $\sim 2 - 10\ \mathrm{deg}^2$ 
for the total area.  These degraded accuracies are still
sufficiently well determined that telescopic searches for MBHB
counterparts have a good chance to be fruitful (although not nearly as
simple as they would be at $z \sim 1$).

In all cases, the GW distance determination is extremely accurate: For
all but the highest masses, $\delta D_L/D_L \lesssim 0.7\%$ at $z =
1$, $\lesssim 4\%$ at $z = 3$, and $\lesssim 5\%$ at $z = 5$.
Distance is determined so precisely that these errors are in fact
irrelevant --- weak gravitational lensing will dominate the distance
error budget for all but the most massive MBHB events.

\begin{figure}[t]
\begin{center}
\includegraphics[scale=0.52]{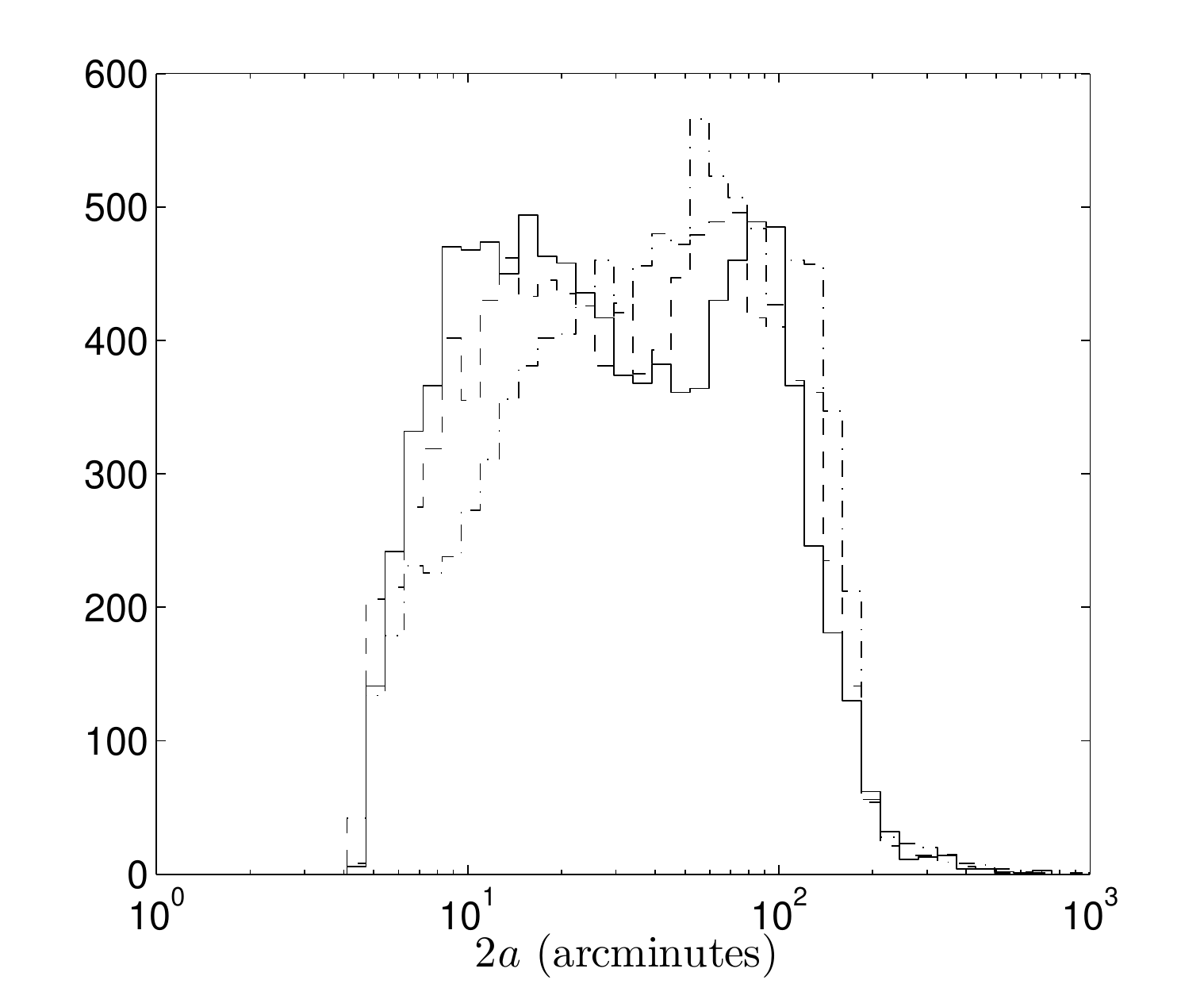}
\includegraphics[scale=0.52]{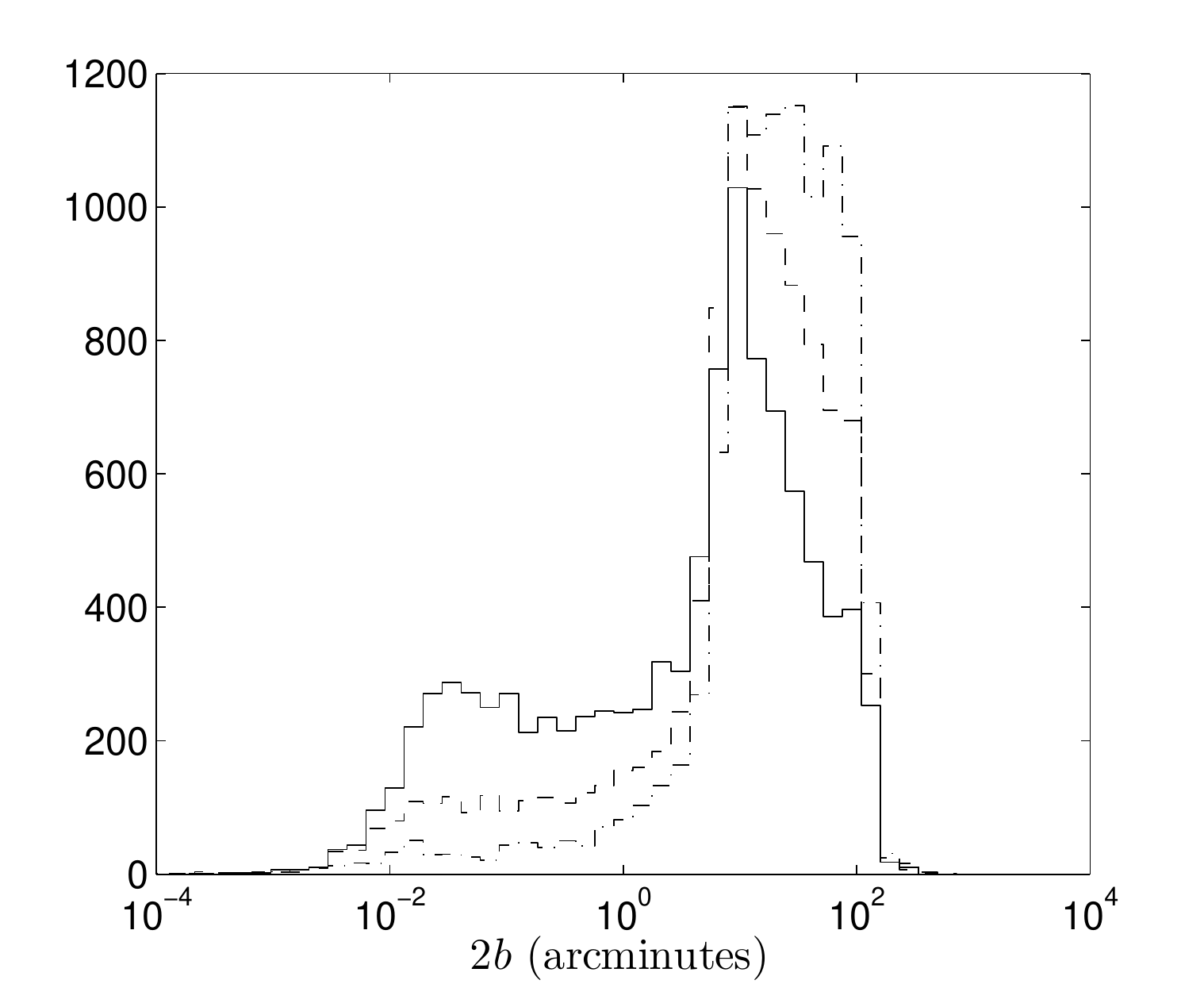}
\caption{Distribution of the major axis $2a$ ({\it left}) and minor axis $2b$
({\it right}) of the sky position error ellipse for $10^4$ binaries with
$m_1 = 10^6 M_\odot$ and $m_2 = 10^5 M_\odot$ ({\it solid line}), $3\times
10^5 M_\odot$ ({\it dashed line}), and $10^6 M_\odot$ 
({\it dash-dotted line}) at
$z = 1$.  Note the bimodal character of the major axis as well as
the long tail down to small minor axis, both of which are particularly prominent for
larger mass ratios.}
\label{fig:final_axes}
\end{center}
\end{figure}

Although the median values reported in Table {\ref{tab:finalvals}}
indicate the typical sky position accuracies we expect, it
should be emphasized that they are taken from broad distributions.
Figure {\ref{fig:final_axes}} presents the distributions we computed
for binaries at $z = 1$ with masses $m_1 = 10^6\,M_\odot$ and $m_2 =
(10^5, 3 \times 10^5, 10^6) M_\odot$.  Note that the major axis
distribution ({\it left}) is rather flat when compared to the minor
axis distribution ({\it right}).  It lacks a single well-defined peak;
in fact, it is actually bimodal for $m_1/m_2 > 1$, with one peak near
$10^\prime - 20^\prime$ and another closer to $1^\circ - 2^\circ$.  We find
that this behavior holds over all mass and redshift cases of interest,
with only slight variations.  At smaller masses, the distribution is
broader than the cases pictured, without strong bimodality; for larger
masses, the distribution is somewhat narrower and tends to develop two
very distinct peaks.  Higher mass ratios tend to accentuate the peaks.
These results hold for higher redshift, except that transitions
between various behaviors occur at smaller total mass [in keeping with
the fact that it is not mass but $(1 + z)$ times the mass that
determines dynamical behavior].

The minor axis distribution exhibits a rather long tail to very small
values, especially when the mass ratio is large.  For $m_1/m_2 = 10$,
the distribution peaks near a minor axis $\sim 10^\prime$, but
extends from roughly $1^{\prime \prime}$ to about $100^\prime$.  As
the mass ratio approaches 1, the peak moves to slightly larger values
(slightly more than $10^\prime$ for $m_1/m_2 = 3$; roughly $30^\prime$ 
for $m_1/m_2 = 1$), and the tail becomes less populated
(although the distributions span roughly the same extent as when
$m_1/m_2 = 10$).  We find that this tail exists for all interesting
mass and redshift combinations, with the same strong dependence on
mass ratio as shown in Figure {\ref{fig:final_axes}}.

Finally, it is worth emphasizing that spin-induced precession has a
significant effect on the accuracy which we report here.  In Paper I,
we compared these results to those obtained with a code which does not
include precession physics.  Position and distance accuracy are
determined only by the detector's motion in this case.  We found that
precession effects reduce the major axis of the sky position error
ellipse by a factor of $\sim 2 - 7$ and the minor axis by a factor of $\sim
2 - 10$.  The distance error is likewise improved by a factor of $\sim 2
- 7$.  Factors of a few or more improvement are also seen when
additional wave harmonics are included {\citep{arunetal,ts08}}.  We
are eager to see whether the improvements from precession and from
higher harmonics can be combined.

\section{Time and location dependence of the GW pixel}
\label{sec:results}

We turn now to a detailed discussion of how well GWs localize an MBHB
system as a function of time before final merger and as a function of
sky location.  We begin by discussing the analysis of K07, 
which presents a clever algorithm for estimating
extrinsic parameter errors as a function of time until merger (although
at present it does not include spin precession).  We demonstrate that
their analysis unfortunately underestimates final position errors by
roughly a factor of $\sim 2 - 4$ or more (in angle) due to neglect of
certain parameter correlations; the underestimate is much less severe
a week or more prior to merger.  We then present our own results for
the time dependence of the GW pixel, using K07 for comparison where
appropriate.  Finally, we conclude with a brief study of the pixel's
dependence on sky location.

\subsection{Summary of K07}
\label{sec:khmf}

K07 have devised a new method, the harmonic mode
decomposition (HMD), to solve for the extrinsic parameter errors as a
function of time to merger.  In the HMD, modulations caused by {\em
LISA}'s motion are decoupled from the much faster inspiral timescales
and are then expanded in a Fourier series.  The resulting expression
for the measured signal features a time-dependent piece with no
dependence on the extrinsic parameters and a time-independent piece
with all the parameter dependence.  As a result, when Monte Carlo
simulations are done across parameter space, the time-dependent
integrals do not need to be recomputed for each sample of the
distribution.  This makes it possible to quickly survey the estimated
parameter errors across a wide range of parameter space.

As already emphasized, the waveform model used by K07 does not (yet)
include the impact of spin precession.  As such, we intend to use
their results as a baseline against which the
impact of spin precession can be compared.  Before doing so, 
we first checked to make sure that their
results were in agreement with a variant of our code which does not
include spin precession {\citep{h02}}.  To our surprise, we found that
the final position accuracy predicted by K07 was typically a factor
$\sim 2$ (in angle) more accurate than our code predicted.

After detailed study of the HMD algorithm and comparison with our
(precession free) code, we believe we understand the primary source of
this disagreement.  K07 define a set of ``slow'' parameters
$\boldsymbol{\theta}_{\mathrm{slow}}$, which correspond (with some
remappings) to our extrinsic parameters: $\ln D_L$, $\cos
\bar{\theta}_{L}$, $\cos \bar{\theta}_N$, $\bar{\phi}_{L}$, and
$\bar{\phi}_N$.  Here $(\bar{\theta}_L, \bar{\phi}_L)$ defines the
orientation of the orbital angular momentum $\mathbf{\hat{L}}$, which
is constant when spin precession is ignored.  K07 also define a set
of ``fast'' parameters $\boldsymbol{\theta}_{\mathrm{fast}}$, which,
modulo the exclusion of spin, map to our intrinsic parameters: $t_c$,
$\Phi_c$, $\ln \mathcal{M}$, and $\ln \eta$.  In their formulation of
the HMD, K07 approximate the cross-correlation between intrinsic and
extrinsic parameters to be zero.  Although the correlations between
intrinsic and extrinsic parameters tend to be small, they are not
zero.  We find that they typically range in magnitude from about 0.1
to 0.4, sometimes reaching $\sim 0.8$.  Neglecting these correlations
altogether leads to a systematic underestimate in extrinsic parameter
errors.

An example of this is shown in Tables {\ref{tab:full_errs_and_corrs}}
and {\ref{tab:khmf_errs_and_corrs}}.  To produce the data shown in
Table {\ref{tab:full_errs_and_corrs}}, we compute the Fisher matrix
$\Gamma_{ab}^{\rm tot}$ and then invert for the covariance matrix,
$\Sigma^{ab} = (\Gamma_{\rm tot}^{-1})^{ab}$.  Table
{\ref{tab:full_errs_and_corrs}} then presents a slightly massaged
representation of this matrix: Diagonal elements are the 1 $\sigma$
errors $(\Sigma^{aa})^{1/2}$, and off-diagonal elements are the
correlation coefficients $c^{ab} =
\Sigma^{ab}(\Sigma^{aa}\Sigma^{bb})^{-1/2}$.  Take particular note of
the magnitude of the correlations between intrinsic and extrinsic
parameters (the upper right-hand portion of Table 
\ref{tab:full_errs_and_corrs}).  Many entries
have values $\sim 0.2 - 0.3$, and two have values $\sim 0.7 - 0.8$.

\begin{table}[p]
\begin{center}
\caption{Full errors and correlations}
\begin{tabular}{cccccc|cccccc}
\tableline \tableline
&
$\ln D_L$ &
$\cos\bar{\theta}_L$ &
$\cos\bar{\theta}_N$ &
$\bar{\phi}_L$ &
$\bar{\phi}_N$ &
$t_c$ &
$\Phi_c$ &
$\ln \mathcal{M}$ &
$\ln\eta$ &
$\beta$ &
$\sigma$ \\

\tableline

$\ln D_L$ & $0.233$ & $-0.984$ & $0.878$ & $0.509$ & $0.213$ & $0.0801$ & $0.246$ & $0.227$ & $-0.186$ & $0.205$ & $-0.106$ \\
$\cos\bar{\theta}_L$ & $\cdots$ & $0.467$ & $-0.861$ & $-0.350$ & $-0.071$ & $-0.040$ & $-0.098$ & $-0.178$ & $0.138$ & $-0.156$ & $0.0622$ \\
$\cos\bar{\theta}_N$ & $\cdots$ & $\cdots$ & $0.0006$ & $0.465$ & $0.203$ & $0.0709$ & $0.231$ & $0.201$ & $-0.166$ & $0.181$ & $-0.095$ \\
$\bar{\phi}_L$ & $\cdots$ & $\cdots$ & $\cdots$ & $0.687$ & $0.782$ & $0.232$ & $0.827$ & $0.337$ & $-0.317$ & $0.328$ & $-0.259$ \\
$\bar{\phi}_N$ & $\cdots$ & $\cdots$ & $\cdots$ & $\cdots$ & $0.0017$ & $0.193$ & $0.691$ & $0.252$ & $-0.244$ & $0.250$ & $-0.210$ \\
\tableline
$t_c$ & $\cdots$ & $\cdots$ & $\cdots$ & $\cdots$ & $\cdots$ & $63.1$ & $0.705$ & $0.923$ & $-0.955$ & $0.942$ & $-0.993$ \\
$\Phi_c$ & $\cdots$ & $\cdots$ & $\cdots$ & $\cdots$ & $\cdots$ & $\cdots$ & $4.00$ & $0.742$ & $-0.747$ & $0.748$ & $-0.726$ \\
$\ln \mathcal{M}$ & $\cdots$ & $\cdots$ & $\cdots$ & $\cdots$ & $\cdots$ & $\cdots$ & $\cdots$ & $0.0010$ & $-0.995$ & $0.998$ & $-0.956$ \\
$\ln\eta$ & $\cdots$ & $\cdots$ & $\cdots$ & $\cdots$ & $\cdots$ & $\cdots$ & $\cdots$ & $\cdots$ & $0.303$ & $-0.999$ & $0.981$ \\
$\beta$ & $\cdots$ & $\cdots$ & $\cdots$ & $\cdots$ & $\cdots$ & $\cdots$ & $\cdots$ & $\cdots$ & $\cdots$ & $1.11$ & $-0.971$ \\
$\sigma$ & $\cdots$ & $\cdots$ & $\cdots$ & $\cdots$ & $\cdots$ & $\cdots$ & $\cdots$ & $\cdots$ & $\cdots$ & $\cdots$ & $0.722$ \\
\tableline
\end{tabular}

\tablecomments{Example of errors (diagonal elements) and correlations
(off-diagonal elements) for a binary with $m_1 = 3 \times 10^6\,M_\odot$ and $m_2
= 10^6\,M_\odot$ at $z = 1$.  The errors in $\bar{\phi}_L$, $\bar{\phi}_N$, and $\Phi_c$ are measured in
radians; the error in $t_c$ is measured in seconds.  This example was taken from the same
Monte Carlo distribution used to make Table
{\ref{tab:compare_full_vs_khmf}}; in this particular case, the
randomly distributed parameters have the values $\cos\bar{\theta}_L = -0.628$, $\cos\bar{\theta}_N = 0.850$, $\bar{\phi}_L =
3.50$ rad, $\bar{\phi}_N = 0.514$ rad, $t_c = 6.90 \times 10^7$ s, $\beta = 1.48$, and $\sigma = 0.107$.  Entries containing
ellipses can be found by symmetry.}
\label{tab:full_errs_and_corrs}
\end{center}
\end{table}

\begin{table}[p]
\begin{center}
\caption{Errors and correlations neglecting correlations between
intrinsic and extrinsic parameters}
\begin{tabular}{cccccc|cccccc}
\tableline \tableline
&
$\ln D_L$ &
$\cos\bar{\theta}_L$ &
$\cos\bar{\theta}_N$ &
$\bar{\phi}_L$ &
$\bar{\phi}_N$ &
$t_c$ &
$\Phi_c$ &
$\ln \mathcal{M}$ &
$\ln \eta$ &
$\beta$ &
$\sigma$ \\
\tableline
$\ln D_L$ & $0.142$ & $-0.999$ & $0.721$ & $0.999$ & $0.0602$ & 0 & 0 & 0 & 0 & 0 & 0 \\
$\cos \bar{\theta}_L$ & $\cdots$ & $0.294$ & $-0.721$ & $-0.999$ & $-0.0611$ & 0 & 0 & 0 & 0 & 0 & 0 \\
$\cos \bar{\theta}_N$ & $\cdots$ & $\cdots$ & $0.0004$ & $0.719$ & $0.0576$ & 0 & 0 & 0 & 0 & 0 & 0 \\
$\bar{\phi}_L$ & $\cdots$ & $\cdots$ & $\cdots$ & $0.120$ & $0.0664$ & 0 & 0 & 0 & 0 & 0 & 0 \\
$\bar{\phi}_N $ & $\cdots$ & $\cdots$ & $\cdots$ & $\cdots$ & $0.0010$ & 0 & 0 & 0 & 0 & 0 & 0 \\
\tableline
$t_c$ & $\cdots$ & $\cdots$ & $\cdots$ & $\cdots$ & $\cdots$ & $61.3$ & $0.990$ & $0.929$ & $-0.959$ & $0.946$ & $-0.993$ \\
$\Phi_c$ & $\cdots$ & $\cdots$ & $\cdots$ & $\cdots$ & $\cdots$ & $\cdots$ & $2.10$ & $0.969$ & $-0.988$ & $0.981$ & $-0.999$ \\
$\ln \mathcal{M}$ & $\cdots$ & $\cdots$ & $\cdots$ & $\cdots$ & $\cdots$ & $\cdots$ & $\cdots$ & $0.0010$ & $-0.995$ & $0.998$ & $-0.960$ \\
$\ln\eta$ & $\cdots$ & $\cdots$ & $\cdots$ & $\cdots$ & $\cdots$ & $\cdots$ & $\cdots$ & $\cdots$ & $0.288$ & $-0.999$ & $0.983$ \\
$\beta$ & $\cdots$ & $\cdots$ & $\cdots$ & $\cdots$ & $\cdots$ & $\cdots$ & $\cdots$ & $\cdots$ & $\cdots$ & $1.05$ & $-0.974$ \\
$\sigma$ & $\cdots$ & $\cdots$ & $\cdots$ & $\cdots$ & $\cdots$ & $\cdots$ & $\cdots$ & $\cdots$ & $\cdots$ & $\cdots$ & $0.697$ \\
\tableline
\end{tabular}

\tablecomments{Example of errors (diagonal elements) and correlations
(off-diagonal elements) for a binary with $m_1 = 3 \times 10^6\,M_\odot$ and $m_2
= 10^6\,M_\odot$ at $z = 1$ if correlations between intrinsic and extrinsic parameters are neglected.  The errors in $\bar{\phi}_L$, $\bar{\phi}_N$, and $\Phi_c$ are measured in
radians; the error in $t_c$ is measured in seconds.  This example was taken from the same
Monte Carlo distribution used to make Table
{\ref{tab:compare_full_vs_khmf}}; in this particular case, the
randomly distributed parameters have the values $\cos\bar{\theta}_L = -0.628$, $\cos\bar{\theta}_N = 0.850$, $\bar{\phi}_L =
3.50$ rad, $\bar{\phi}_N = 0.514$ rad, $t_c = 6.90 \times 10^7$ s, $\beta = 1.48$, and $\sigma = 0.107$.  Entries containing
ellipses can be found by symmetry.}
\label{tab:khmf_errs_and_corrs}
\end{center}
\end{table}

To see what effect neglecting the intrinsic-extrinsic correlations
has, we repeat this exercise, with a slight modification: We compute
$\Gamma^{\rm tot}_{ab}$ as before, but we now set to zero entries
corresponding to mixed intrinsic/extrinsic parameters.  For example,
we set by hand $\Gamma_{\ln D_L,\ \ln \mathcal {M}} = 0$.  We then
invert this matrix to obtain $\Sigma^{ab}$.  The result is shown in
Table {\ref{tab:khmf_errs_and_corrs}}.  Note that mean parameter
error (diagonal entries) is often significantly smaller than errors
when these correlations are not ignored.  The impact of correlations
between intrinsic and extrinsic parameters is clearly not negligible.

Table {\ref{tab:compare_full_vs_khmf}} gives further examples
illustrating the impact of neglecting these correlations on our
estimates of {\it LISA}'s localization accuracy.  We show 10 points
drawn from a $10^4$ binary Monte Carlo run; all use the same masses
and redshifts ($m_1 = 3 \times 10^6\,M_\odot$, $m_2 = 10^6\,M_\odot$, and
$z = 1$) but have different (randomly distributed) sky positions,
orientations, spins, and $t_c$.  For these parameters, we find that
neglecting intrinsic-extrinsic correlations causes one to
underestimate the major axis of the position ellipse by a (median)
factor of $\sim 2$ and the minor axis by a factor of $\sim 3-4$; the area is
underestimated by a factor of $\sim 6-7$.

\begin{table}[t]
\begin{center}
\caption{Example sky position error measures from Monte Carlo sample, comparing full Fisher matrix technique with the K07 approximation}
\begin{tabular}{cccccccc}
\\
\tableline \tableline
\multicolumn{2}{c}{$2a$ (arcmin)} & 
& 
\multicolumn{2}{c}{$2b$ (arcmin)} &
& 
\multicolumn{2}{c}{$\Delta \Omega_N\ (\mathrm{deg}^2)$} \\
\cline{1-2} \cline {4-5} \cline{7-8} 
Full &
K07 &
& 
Full &
K07 &
& 
Full &
K07 \\

\tableline

$201$  & $63.5$ & & $59.0$ & $15.1$ & & $2.59$  & $0.210$ \\
$165$  & $120$  & & $108$  & $88.6$ & & $3.90$  & $2.32$  \\
$117$  & $61.6$ & & $81.0$ & $14.3$ & & $2.07$  & $0.193$ \\
$197$  & $69.7$ & & $46.7$ & $13.4$ & & $2.01$  & $0.204$ \\
$10.9$ & $7.23$ & & $8.21$ & $5.03$ & & $0.0196$ & $0.00793$ \\
$197$  & $51.2$ & & $55.2$ & $13.3$ & & $2.37$  & $0.149$ \\
$46.7$ & $26.8$ & & $36.5$ & $9.19$ & & $0.372$ & $0.0538$ \\
$18.1$ & $12.7$ & & $15.0$ & $6.37$ & & $0.0595$ & $0.0177$ \\
$155$  & $92.4$ & & $88.5$ & $16.2$ & & $2.98$  & $0.326$ \\
$146$  & $143$  & & $139$  & $10.9$ & & $4.43$  & $0.342$ \\
\tableline
\end{tabular}

\tablecomments{Ten Monte Carlo points comparing sky position accuracy
for binaries with $m_1 = 3 \times 10^6\,M_\odot$ and $m_2 = 10^6\,M_\odot$
at $z = 1$.}
\label{tab:compare_full_vs_khmf}
\end{center}
\end{table}

Our conclusion is that the HMD technique developed by K07 is overly
optimistic by a factor of $\sim 2-4$ or more (in angle) regarding the
final accuracy with which GWs can locate an MBHB event on the sky.  As
a prelude to the time evolution study we present in \S\
{\ref{sec:timeevolve}}, we also examined how this underestimate
evolves as merger is approached.  To our relief, it appears that this
underestimate is {\it much} smaller prior to merger: For the handful
of cases we examined, the factor of $2 - 4$ underestimate in angle
falls to a mere $10\% - 25\%$ offset one week prior to merger.  We find
that the offset plateaus at this level, remaining at a few tens of percent
up to 28 days before merger.

Accounting for this systematic final underestimate, we thus find
K07's results to be a good baseline against which to compare our
results.  This comparison makes it possible to assess the extent to
which spin precession improves our ability to locate massive black
hole binaries prior to the final merger.

\subsection{Results I: Time evolution of localization accuracy}
\label{sec:timeevolve}

We finally come to the main results of this paper, the time evolution
of our ability to localize MBHB systems using GWs when spin precession
is included.  The results summarized in \S\ {\ref{sec:review}}
describe the size of the GW pixel (sky position error ellipse and
luminosity distance) at the end of inspiral.  We do not at this time
incorporate any information regarding the merger and ringdown phases.  
The end-of-inspiral\footnote{Throughout this
paper, ``final'' accuracy refers to the end of inspiral.} localization
accuracies are good enough that searching the GW pixel for
counterparts to MBHB coalescences is likely to be fruitful.  However,
given how little is understood about electromagnetic counterparts to
these events, it is unclear if waiting until these final moments is
the best strategy for such a search.  It will surely be desirable to
also monitor the best-guess location some days or weeks in advance
for electromagnetic precursors to the final merger.  The rate at which
the GW pixel evolves as we approach the merger will have strong
implications for determining the rate at which {\it LISA} data is sent
to the ground.

To this end, we now examine the time dependence of the {\em LISA}
pixel.  We have modified the code from Paper I to stop the calculation
at a specified time before the fiducial merge frequency, equation
\eqref{eq:fmerge}.  We begin the evolution of each binary at the
moment it enters the {\it LISA} band (which we take to occur at
$f_{\rm low} = 3 \times 10^{-5}\,{\rm Hz}$).  Because we randomly
distribute $t_c$ over our (assumed) 3 yr {\it LISA} mission, some
sources are already in band at the mission's start; consequently,
these sources begin at $f > f_{\rm low}$.  The binary's evolution is
then followed until it reaches a GW frequency $f_{\mathrm{stop}} =
f(t(f_{\mathrm{merge}})-N)$, where $N$ is the number of days before
merger that we want to stop the signal.  (Choosing $N = 0$ duplicates
the analysis of Paper I.)

\begin{figure}[!ht]
\begin{center}
\includegraphics[scale=0.9]{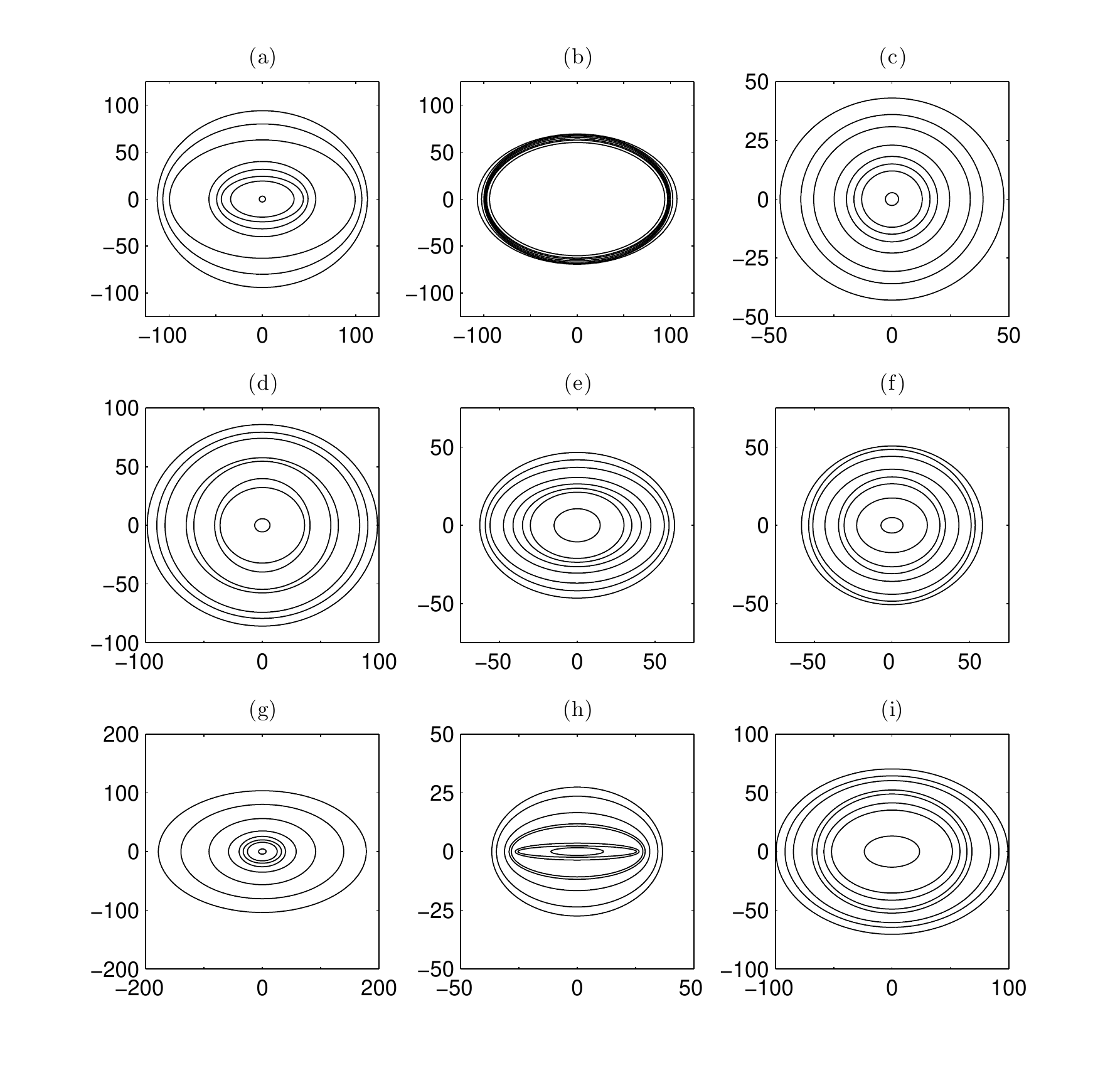}
\caption{Evolution of the sky position error ellipse for nine
individual binaries selected from a set of $10^4$.  All have $m_1 =
10^6 M_\odot$, $m_2 = 3\times 10^5 M_\odot$, and $z = 1$.  The
ellipses are oriented so their major axes are parallel to the $x$-axis
and their minor axes are parallel to the $y$-axis; the axes are
labeled in arcminutes.  From outside in, the ellipses are evaluated at
28, 21, 14, 7, 4, 2, 1, and 0 days before merger.}
\label{fig:pixels}
\end{center}
\end{figure}

Figure \ref{fig:pixels} shows the error ellipse evolution for nine
examples taken from a sample of $10^4$ computed with $m_1 = 10^6
M_\sun$, $m_2 = 3\times 10^5 M_\sun$ and $z = 1$.  We show results for
$N = $ 0, 1, 2, 4, 7, 14, 21, and 28.  For each binary, the major axis
is plotted on the $x$-axis, while the minor axis is plotted on the
$y$-axis; we do not show how each ellipse would be oriented on the
sky.  The results shown in Figures \ref{fig:pixels}{\it a} and 
\ref{fig:pixels}{\it b} were selected by hand
from the distribution as examples of contrasting behavior.  The binary 
in Figure \ref{fig:pixels}{\it a}
shows a dramatic change in the error ellipse with time, especially in
the last day before merger.  In that day, the binary is localized to
an ellipse with $2a = 6.67^\prime$ and $2b = 6.25^\prime$, an area
$\sim 60$ times smaller than at $N = 1$.  By contrast, 
the binary in Figure \ref{fig:pixels}{\it b}
shows almost no change in the error ellipse over the entire four weeks
prior to merger.

These are clearly extreme cases.  Other extremes exist, including
binaries with a minor axis orders of magnitude smaller than the major
axis (see the tail in Fig.\
\ref{fig:final_axes}, {\it right}), binaries where the evolution of one or both
axes is not strictly monotonic, and binaries which have very large
ellipses (essentially filling the sky) for large $N$.  (Such cases
correspond to binaries which are already well into the {\it LISA} band
when the mission starts; for large $N$, there is little baseline for
the various modulations to encode their position.)  To get a sense of
more typical behavior, we selected the binaries in Figures 
\ref{fig:pixels}{\it c} -- \ref{fig:pixels}{\it i} randomly
from the distribution.  There does not appear to be any ``typical''
evolution; each binary exhibits some unique features.  Most binaries,
however, seem to share with the binary in Figure 
\ref{fig:pixels}{\it a} the property that the final day
before merger gives much more information on the position than any day
before it (albeit to a lesser degree).  We will see below that this
feature holds for the medians of almost all mass and redshift cases.
It is worth noting that although K07 agree with us on most of the
other qualitative features of the time dependence, they do not see the
dramatic change in the final day of inspiral.  As we will discuss in
more detail below, this dramatic improvement toward the end of
inspiral is due to spin precession physics.  This is in good agreement
with the expectations of N. Cornish (2005, unpublished) and K07 that 
spin precession
would most dramatically impact the last week or so of inspiral.

\begin{figure}[htb]
\begin{center}
\includegraphics[scale=0.49]{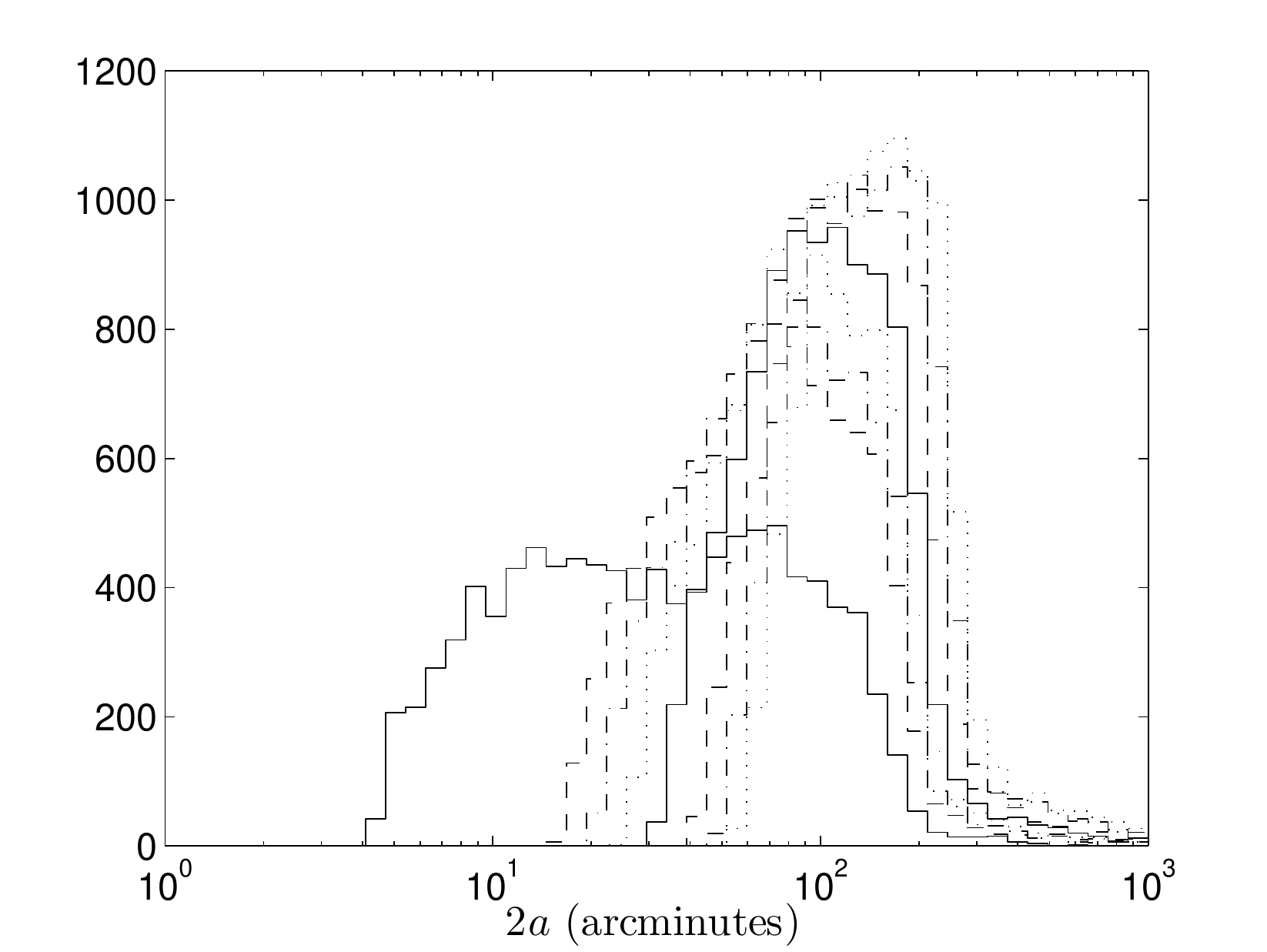}
\includegraphics[scale=0.49]{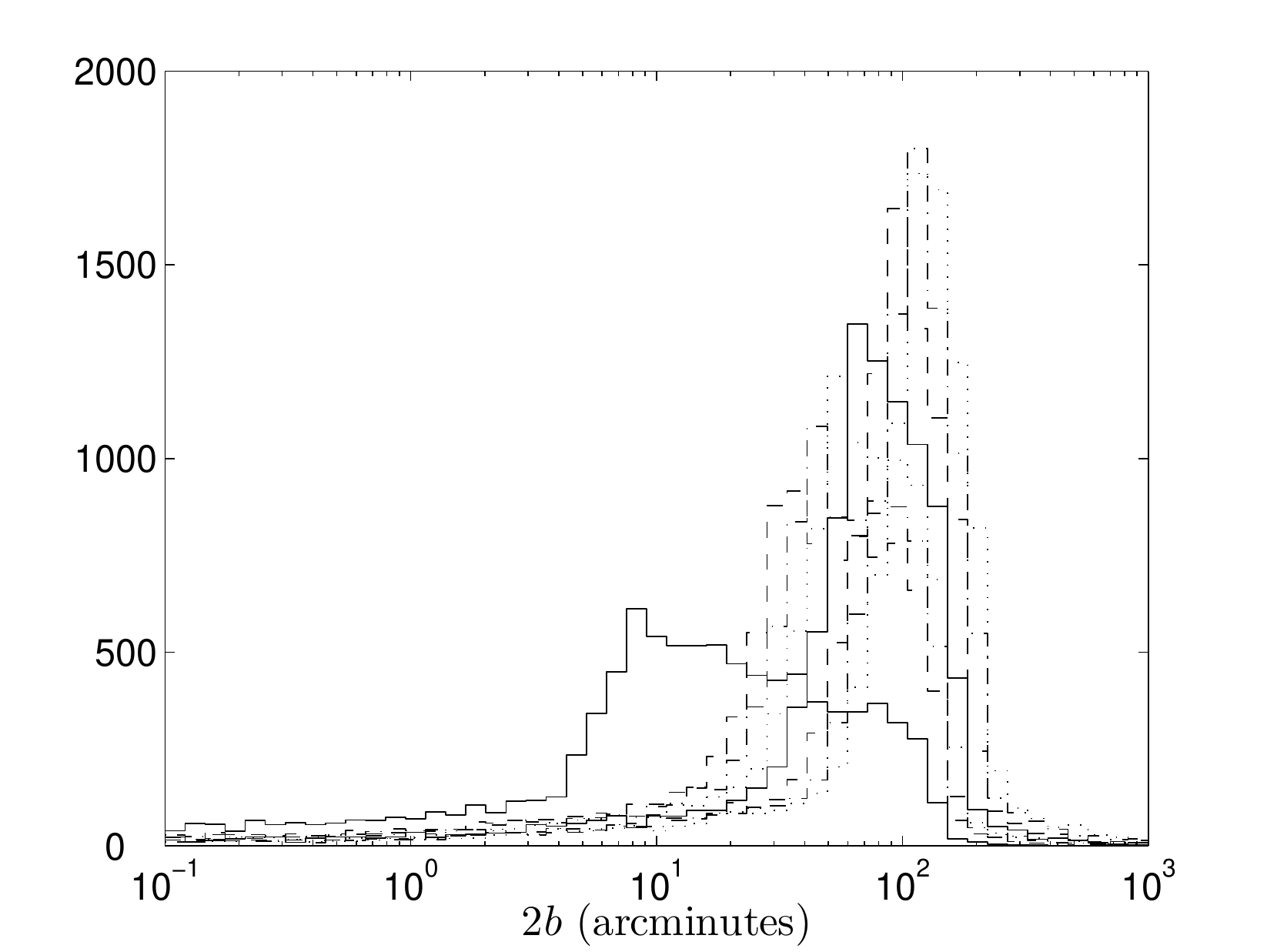}
\caption{Same as Fig.\ \ref{fig:final_axes}, but with 
$m_1 = 10^6 M_\odot$, $m_2 = 3 \times 10^5 M_\odot$, and $z = 1$ at
different values of $N$ (the number of days before merger).  Reading
from left to right, $N = 0$ ({\it solid line}), 1 ({\it dashed line}), 
2 ({\it dash-dotted line}), 4 ({\it dotted line}), 
7 ({\it solid line again}), 14 ({\it dashed line}), 
21 ({\it dash-dotted line}), and 28 ({\it dotted line}).  
Clearly, the largest change --- in
shape as well as median --- happens between merger and one day before.}
\label{fig:axes_evol}
\end{center}
\end{figure}

While the evolution of parameter errors for individual binaries is
interesting, of more relevance is the evolution of the errors'
distribution.  The left panel of Figure \ref{fig:axes_evol} shows
the time dependence of the distribution of the major axis $2a$ for our
model system of $m_1 = 10^6 M_\odot$, $m_2 = 3\times 10^5 M_\odot$,
and $z = 1$.  We can clearly see the evolution to smaller major axis
as the binary nears merger.  Four weeks before merger, the median
major axis is $4.8$ times larger than at merger; this number shrinks
to $3.9$ two weeks before merger, $3.2$ one week before, and $2.5$ two
days before.  As expected from the individual binaries, the most
dramatic change in the distribution occurs during the last day before
merger.  Not only is the median substantially reduced (by a factor of
$2.2$), but the shape sharply changes.  For $N > 0$, the distribution
is distinctly peaked, becoming gradually flatter as $N$ gets smaller.
Over the last day of inspiral, the distribution evolves into the
almost entirely flat, slightly bimodal shape first seen in
Figure \ref{fig:final_axes}.  We find that this same behavior holds for
all masses and redshift cases of interest: A sharply peaked
distribution at $N > 0$ evolves into the flatter, sometimes bimodal
final distributions described in \S\ \ref{sec:review}.  As the total
mass increases, however, the final distributions become so narrow that
the shape change is no longer very clear.  As we might expect, this
transition occurs at smaller total mass for higher $z$.

The right panel of Figure \ref{fig:axes_evol} shows the evolution
of the minor axis $2b$.  Again the distribution slowly changes shape
over time, with the most drastic change occurring in the last day.
Here the final distribution still retains a slight peak, along with
the previously discussed long tail of small errors.  Interestingly,
this tail is present to some degree throughout the evolution.  As the
total mass increases, the final distribution moves to the right until,
as with $2a$, the sharp change of shape disappears.  The same
evolution occurs at higher $z$, again with a shift in the mass scale.

\begin{figure}[!p]
\includegraphics[scale=0.49]{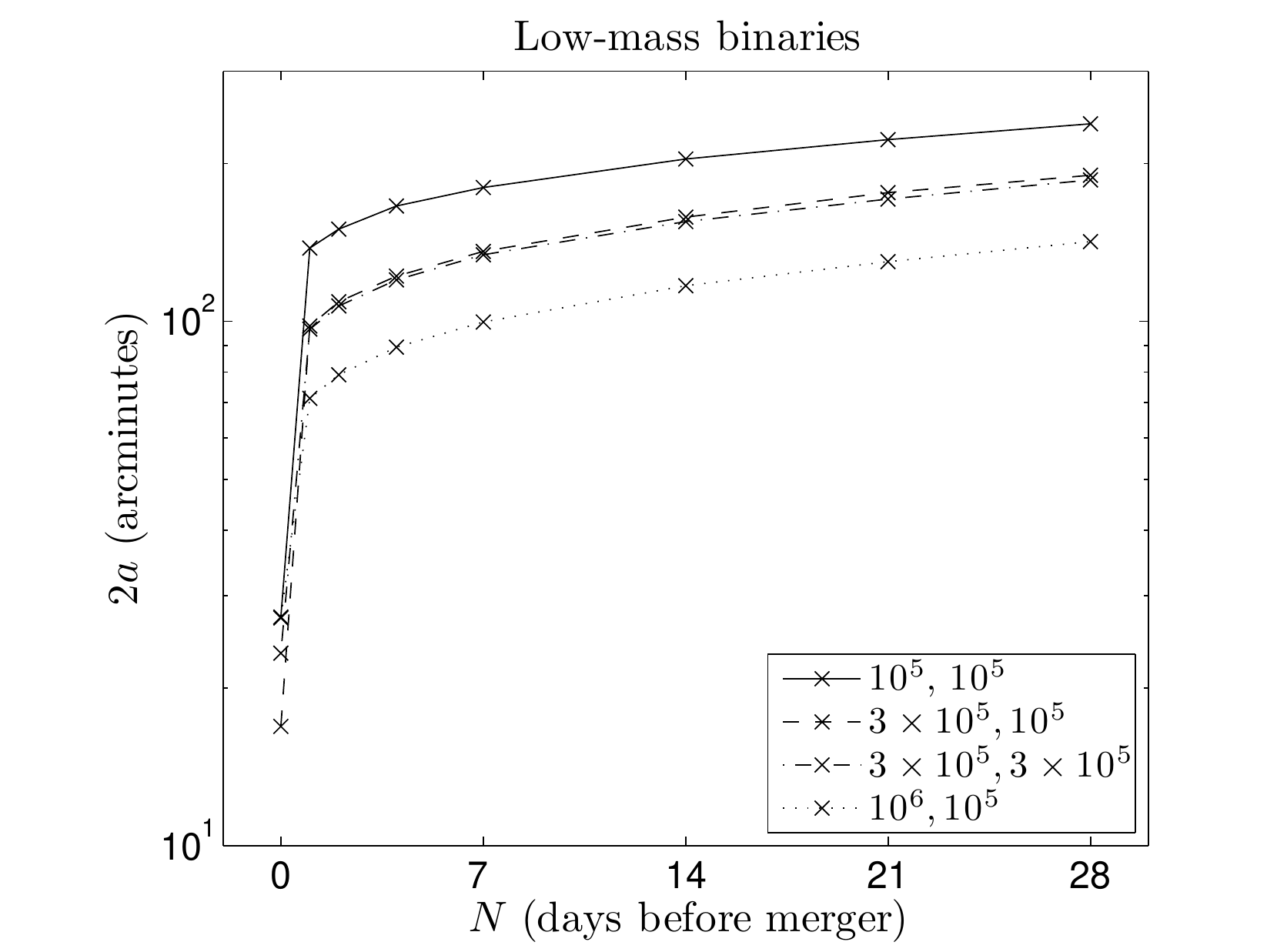}
\includegraphics[scale=0.49]{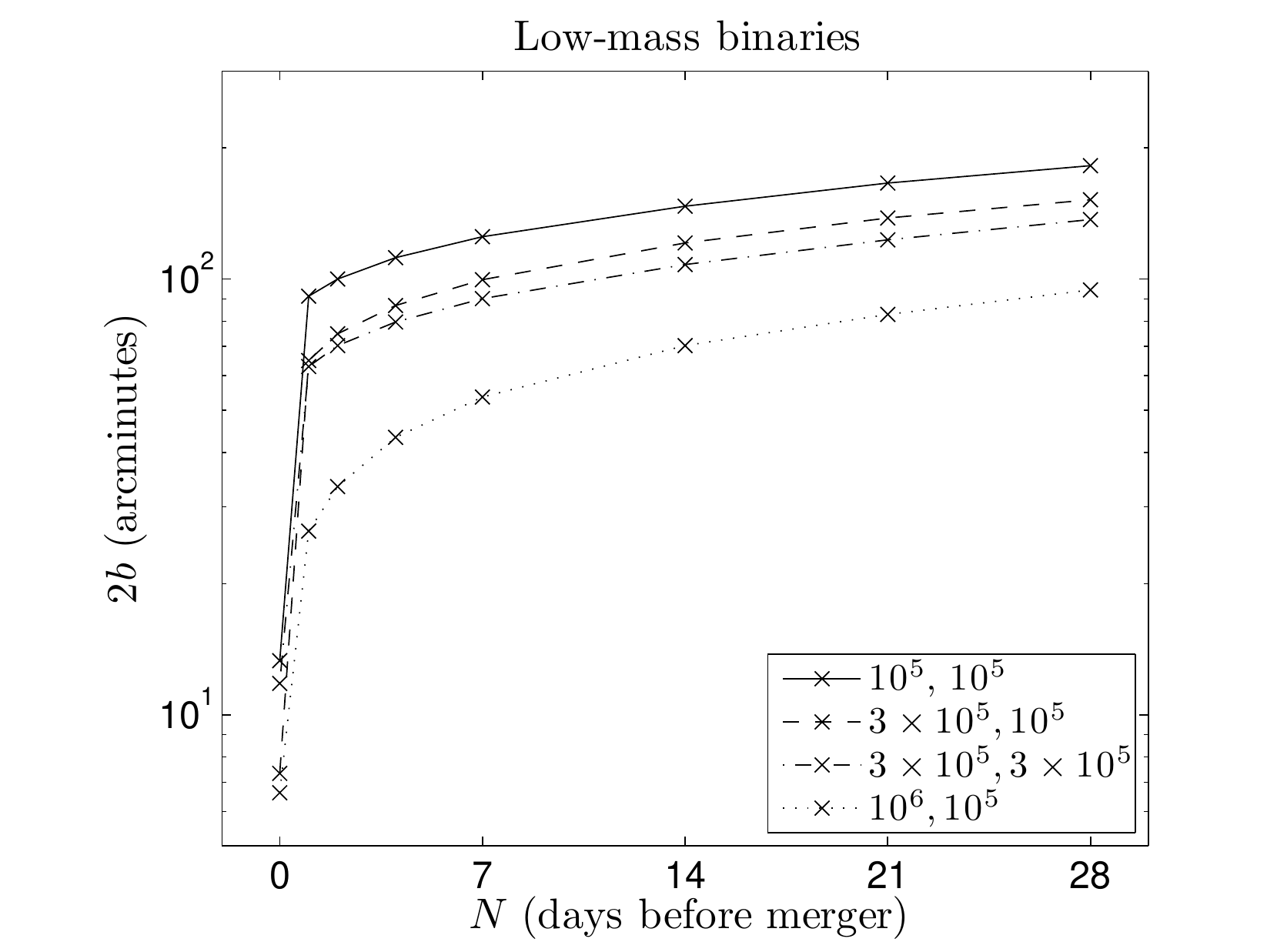}
\includegraphics[scale=0.49]{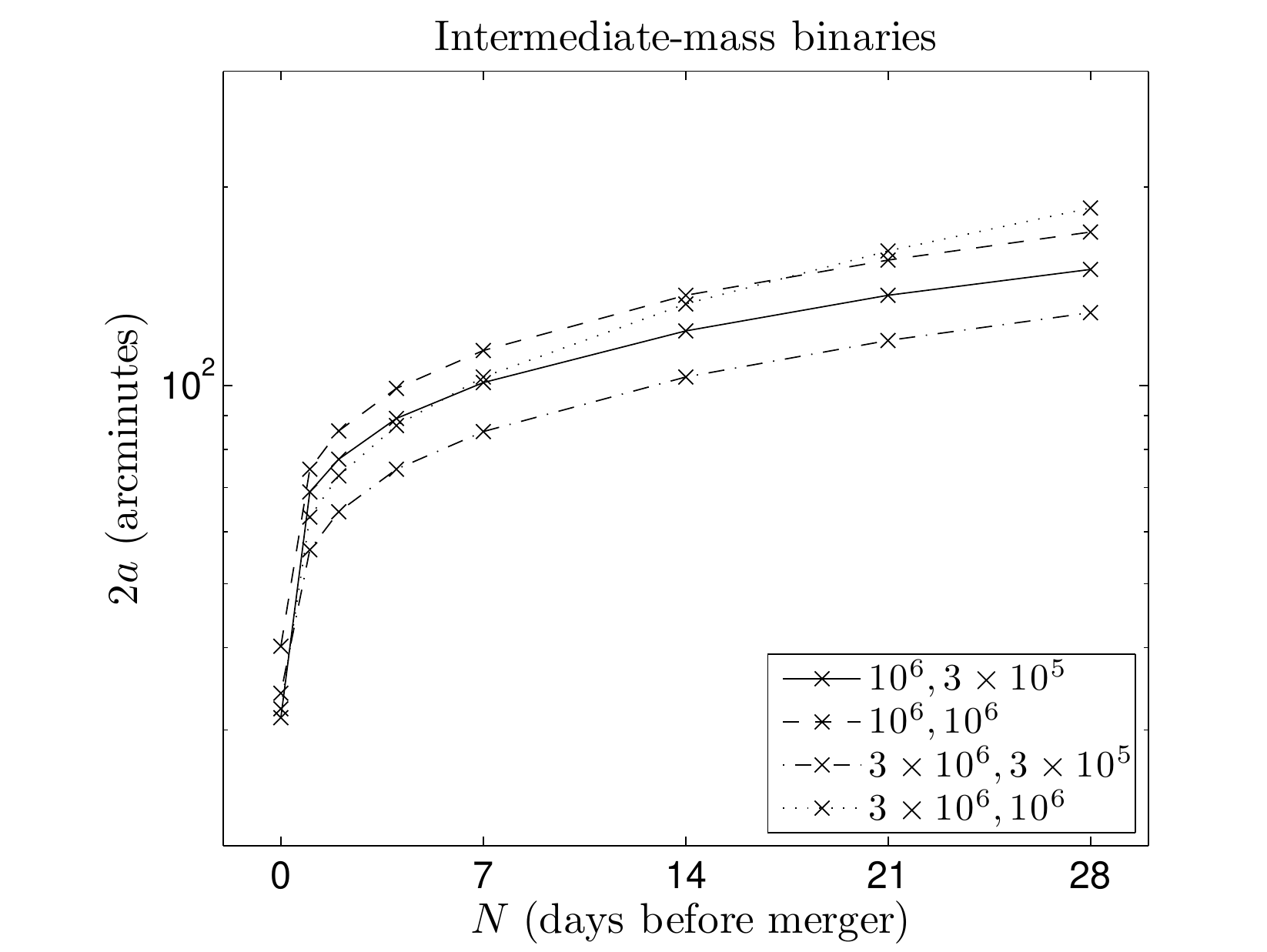}
\includegraphics[scale=0.49]{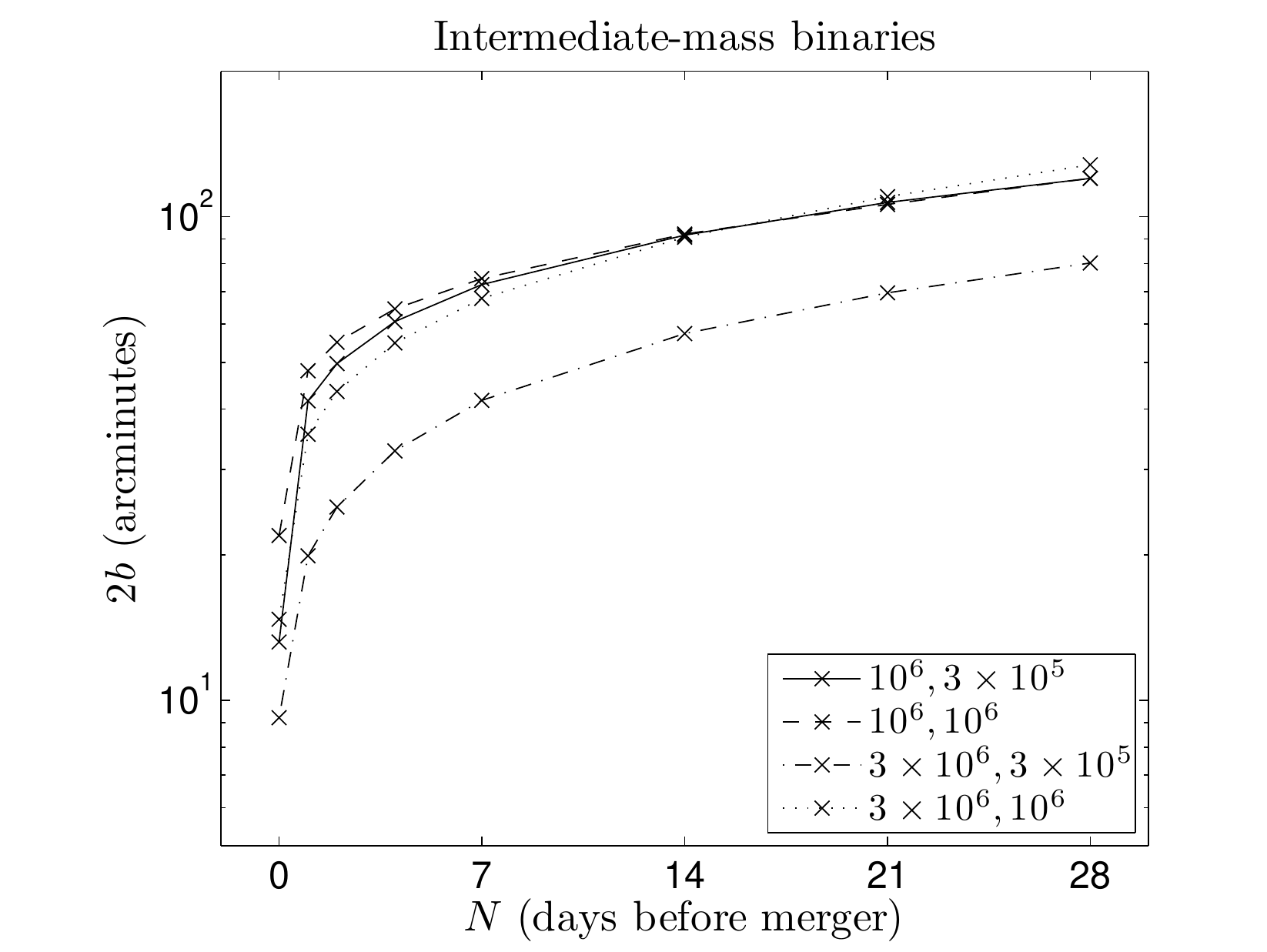}
\includegraphics[scale=0.49]{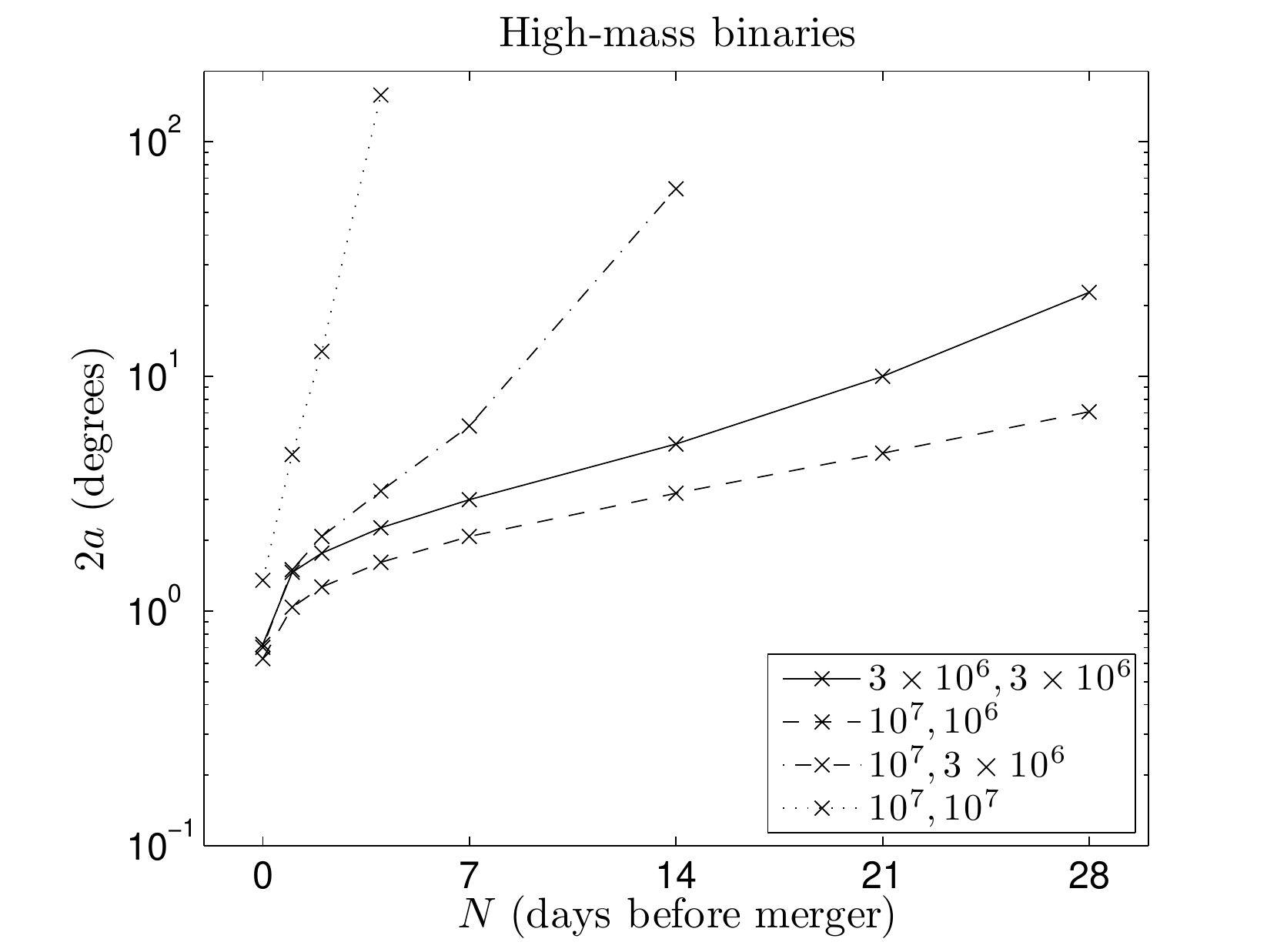}
\includegraphics[scale=0.49]{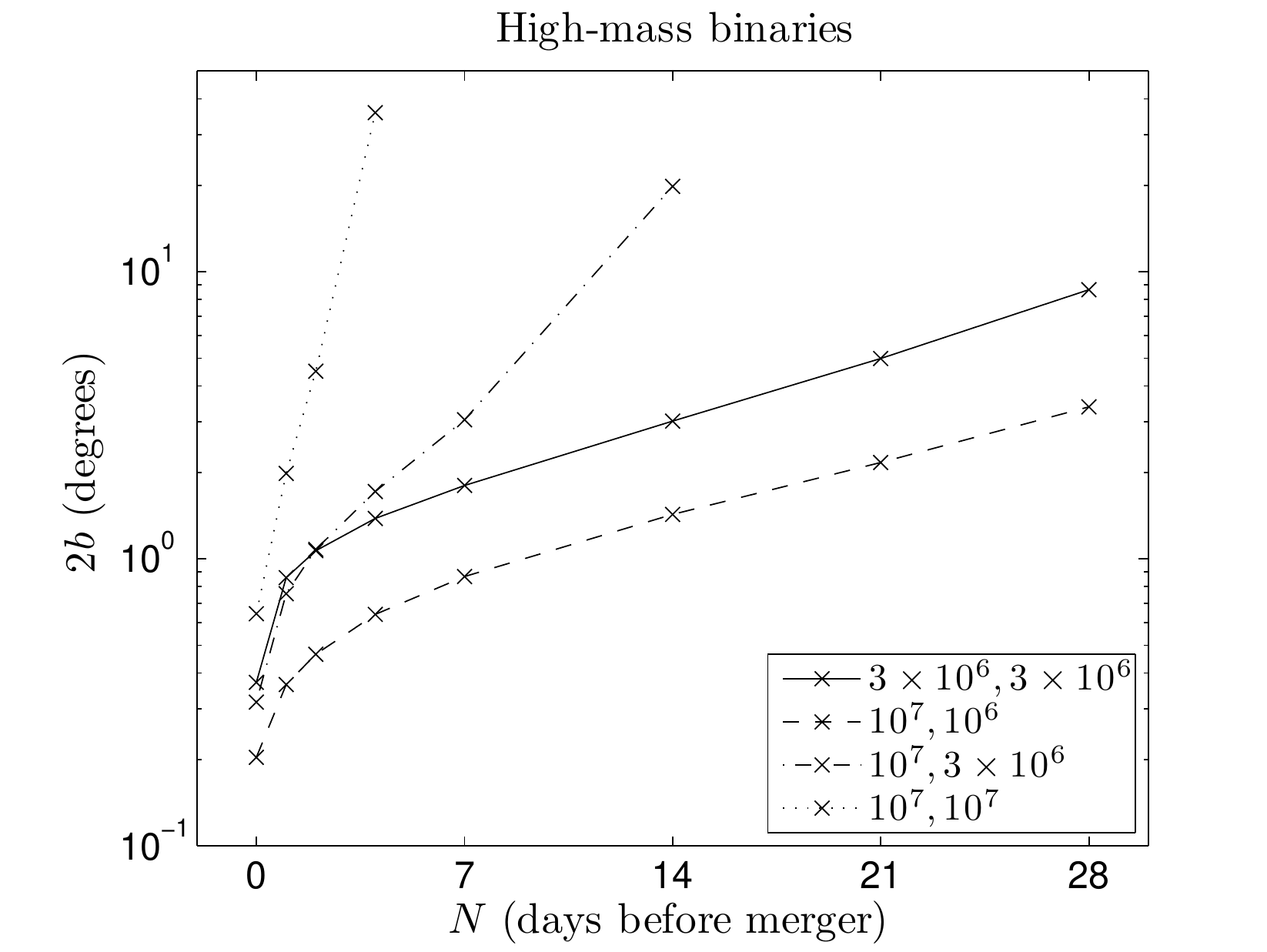}
\caption{Medians of the sky position ellipse axes for Monte Carlo runs
of $10^4$ binaries as a function of time before merger.  Major axes
$2a$ are on the left; minor axes $2b$ are on the right.  Data were only
outputted at the marked points; the lines are there just to guide the
eye.  The masses have been subdivided into ``low,'' ``intermediate,'' and
``high'' groups; the exact values (in units of solar masses) are given in
the legends.  Note also the different scales (arcminutes and degrees)
on the $y$-axis.}
\label{fig:ellipsemedians} 
\end{figure}

To further understand how the error ellipse evolves, consider Figure
{\ref{fig:ellipsemedians}}.  Here we show the evolution of the median
axes for a wide range of masses (``low,'' ``intermediate,'' and
``high'') at $z = 1$.  As in Figures {\ref{fig:pixels}} and
{\ref{fig:axes_evol}}, our calculations produce output for $N = $ 0,
1, 2, 4, 7, 14, 21, and 28; they are connected by lines only to guide
the eye.  Almost all the cases we present show similar behavior: The
ellipses gradually shrink with time before sharply decreasing in size
during the final day.  Significant deviations from this behavior come
from the high-mass binaries, which evolve more drastically at large
$N$ before settling in to resemble the lower mass curves.  This
high-mass deviation is an artifact of our choice of maximum $N$.
Smaller binaries may spend many months or even years in the {\it LISA}
band before merger.  In those cases, enough signal has already been
measured at $N = 28$ to locate the binary reasonably well.  By
contrast, the high-mass binaries spend much less time in
band\footnote{The two highest mass binaries in Fig.\ \ref{fig:ellipsemedians}
are not even
in band a full 28 days.} and have not been measured so well by $N =
28$.  They have to ``catch up'' to the smaller mass binaries over the
first few weeks of our measurement window.  Nearly identical results
were found by K07: Figure 2 of K07 plots the evolution of
sky position for an intermediate-mass binary from $N \simeq 300$.
They find that the measurement accuracy rapidly evolves early in the
measurement, with slopes of angular error versus time very similar to
what we show in Figure {\ref{fig:ellipsemedians}} for high-mass
binaries.

Although these curves are qualitatively quite similar, there are
significant quantitative differences.  For example, the evolution of
the intermediate-mass binaries is less than that of the low-mass
binaries, especially in the last day.  The intermediate-mass major
axes shrink by a factor of $\sim 4-6$ over the entire four-week period,
whereas the low-mass axes shrink by a factor of $\sim 5-11$.  Another important
quantitative difference can be seen by comparing the major and minor
axes.  We find that the ratio $2a/2b$ grows with time in most cases,
indicating that the minor axis tends to shrink more rapidly than the
major axis.  The only exceptions are the previously described
high-mass cases, in which $2a/2b$ shrinks during most or all of the
inspiral.  Presumably the same behavior would also be seen for
lower mass binaries at higher values of $N$; this conclusion is
supported by Figure 4 of {\cite{khm07}}.  For all other cases, $2a/2b$
$\sim 1.3-1.7$ at $N = 28$ and increases to $\sim 1.8-4$ at $N =
0$.  The largest increase is typically in the final day.

What causes this dramatic improvement in the last day of inspiral?
Three factors primarily contribute to our ability to localize a source
on the sky: modulations due to {\it LISA}'s orbital motion,
modulations due to spin precession, and S/N accumulated over time.
Since {\it LISA} moves the same amount in the final day as in any
other day, orbital-induced modulations cannot be the cause.  A great
deal of S/N is accumulated in the last day (typically increasing by a
factor of $\sim 2$ or more), and many parameter errors scale as
(S/N)$^{-1}$.  However, K07 demonstrate that sky position and distance
errors do not scale as (S/N)$^{-1}$ in the last few weeks before merger.
Our ``no precession'' code supports their conclusion: The final jump
in S/N cannot make up for the lack of orbital modulation over such a
short timescale.

\begin{figure}[htb]
\includegraphics[scale=0.49]{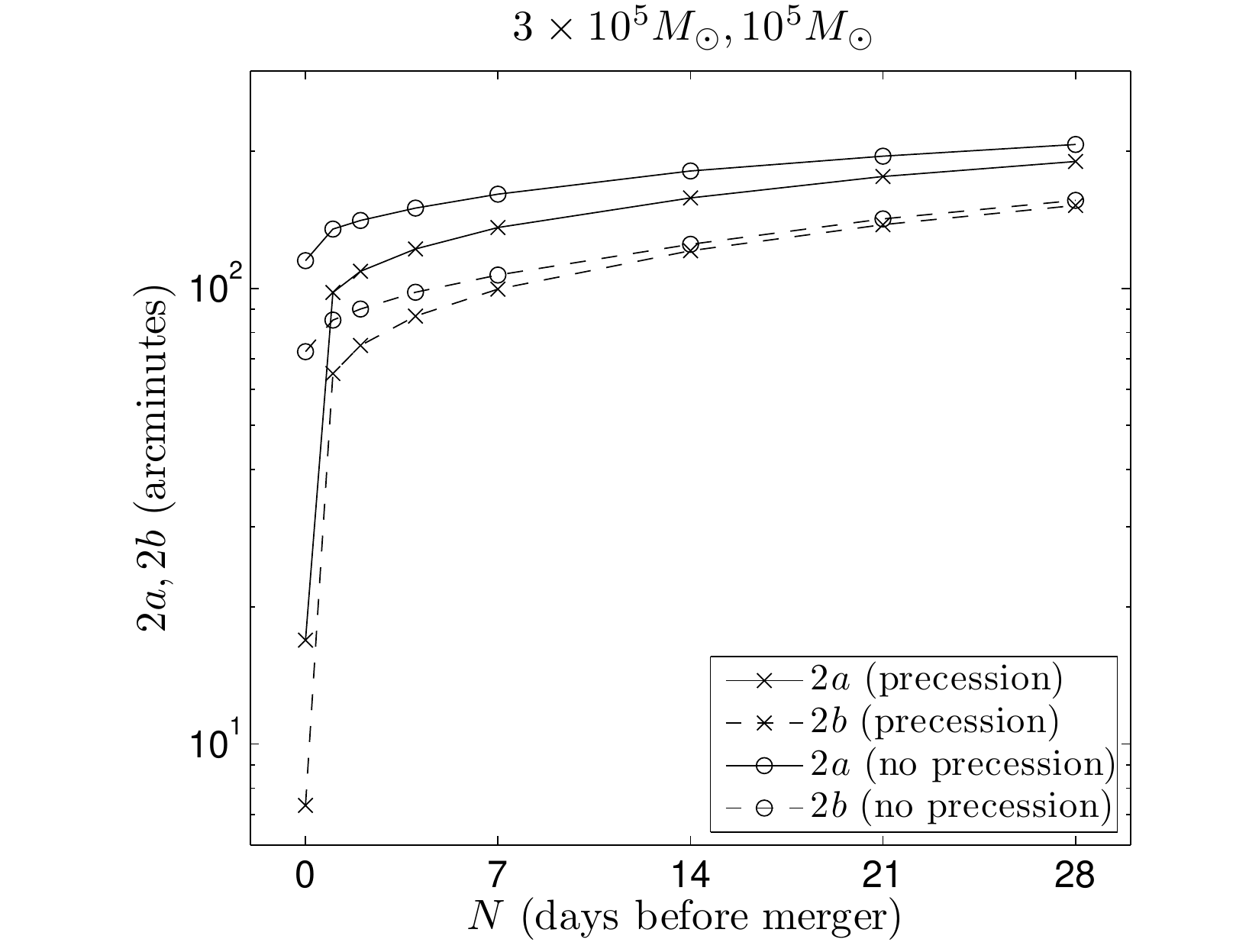}
\includegraphics[scale=0.49]{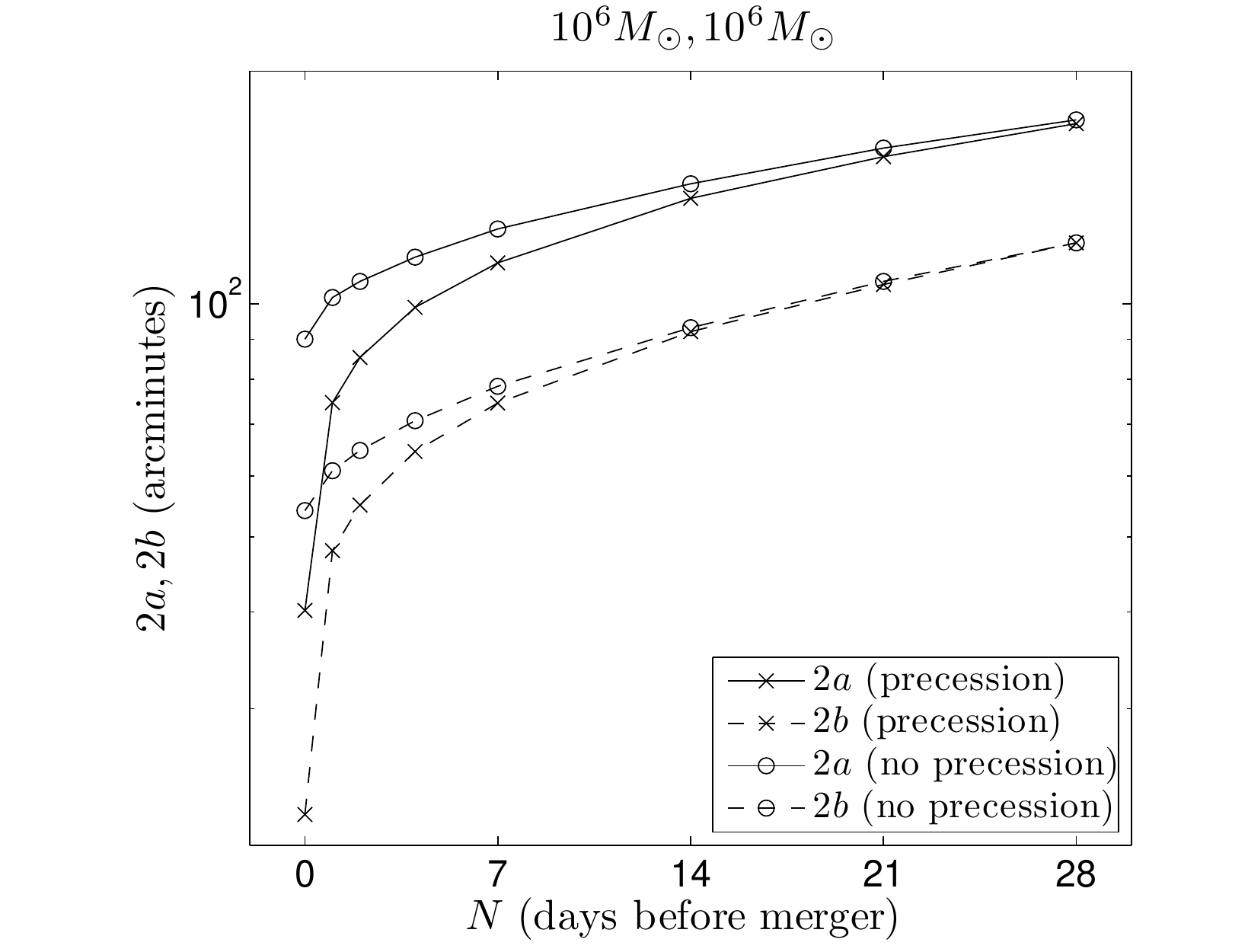}
\caption{Medians of $2a$ and $2b$ as a function of time, comparing an
analysis that accounts for spin-induced precession to one that
neglects it.  Solid lines trace the evolution of $2a$, dashed lines
trace $2b$.  Precession results are marked with crosses, no
precession with circles.  The left plot shows a low-mass case,
$m_1 = 3\times 10^5 M_\odot$ and $m_2 = 10^5 M_\odot$; the right plot
shows an intermediate-mass case, $m_1 = m_2 = 10^6 M_\odot$.  Both
plots are for $z = 1$.}
\label{fig:NPcomparison}
\end{figure}

The remaining possibility is spin-induced precession.  Indeed, we
found in Paper I that the number of modulations due to spin precession
increases dramatically as the binary approaches merger.  This suggests
that the improvement we see is due to the impact of precession.  To
examine this hypothesis, we plot the ``precession'' and ``no
precession'' results together on the same axes.  Figure
\ref{fig:NPcomparison} shows such a plot for a low-mass system and an
intermediate-mass system at $z = 1$.  We see that at $N = 28$ days,
the two codes give similar results for localization accuracy.  Their
predictions gradually diverge as merger is approached.  The greatest
jump between the two codes occurs on the last day before merger,
agreeing with our expectation that precession effects are maximal
then.  The effect is greater in the low-mass case than in the
intermediate-mass case.  Similarly, the ratio $2a/2b$ starts about the
same in both codes, growing very slowly until $N = 1$.  At this point,
it jumps dramatically in the precession code, while staying
roughly the same in the no precession code.  Interestingly, one
effect of precession is that the localization errors track (S/N)$^{-1}$
rather closely.  By breaking the various correlations which made them
deviate from (S/N)$^{-1}$, precession-induced modulations allow the
errors to evolve in a manner that is more consistent with our naive
expectations.

\begin{figure}[!ht]
\begin{center}
\includegraphics[scale=0.52]{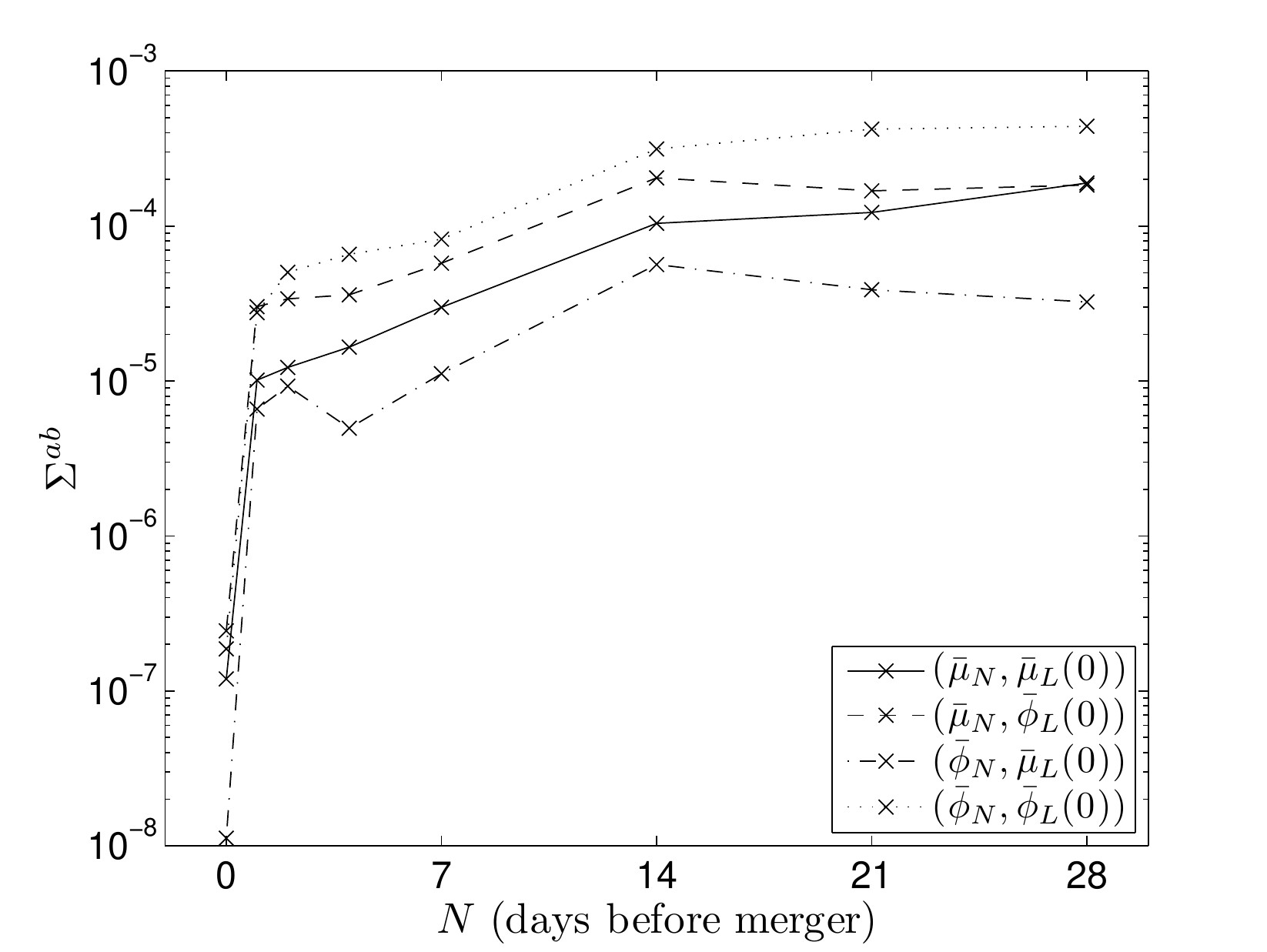}
\caption{Off-diagonal covariance matrix entries illustrating
correlation between sky position and binary orientation as a function
of time, for the binary in Fig.\ \ref{fig:pixels}{\it a}.  The correlations
decrease rapidly in the final day before merger, when precession
effects are maximal.}
\label{fig:muNphiNcovar}
\end{center}
\end{figure}

The time evolution of the correlations of interest, those between the
sky position and binary orientation, are illustrated in Figure
\ref{fig:muNphiNcovar}.  Here we show the off-diagonal components of
the covariance matrix $\Sigma^{ab}$ (where $a \in
\{\bar{\mu}_N,\bar{\phi}_N\}$ and $b \in
\{\bar{\mu}_L(0),\bar{\phi}_L(0)\}$) for the binary in Figure
\ref{fig:pixels}{\it a}.  We found that examining the normalized correlation
coefficients $c^{ab} = \Sigma^{ab}(\Sigma^{aa}\Sigma^{bb})^{-1/2}$ can
mislead since $\Sigma^{aa}$ and $\Sigma^{bb}$ are rapidly evolving at
the same time as $\Sigma^{ab}$.  At large $N$, the two sets of angles
are relatively strongly correlated due to degeneracies in the measured
waveform (\ref{eq:freqdomainsignal}). However, in the last day before
merger, the correlations sharply decrease as precession effects
accumulate.  The reduction of these correlations coincides with the
sudden drop in parameter errors seen in Figure \ref{fig:pixels} (and,
by extension to the entire Monte Carlo run, Figs.\ \ref{fig:axes_evol}
-- \ref{fig:NPcomparison}).

\begin{figure}[!ht]
\begin{center}
\includegraphics[scale=0.52]{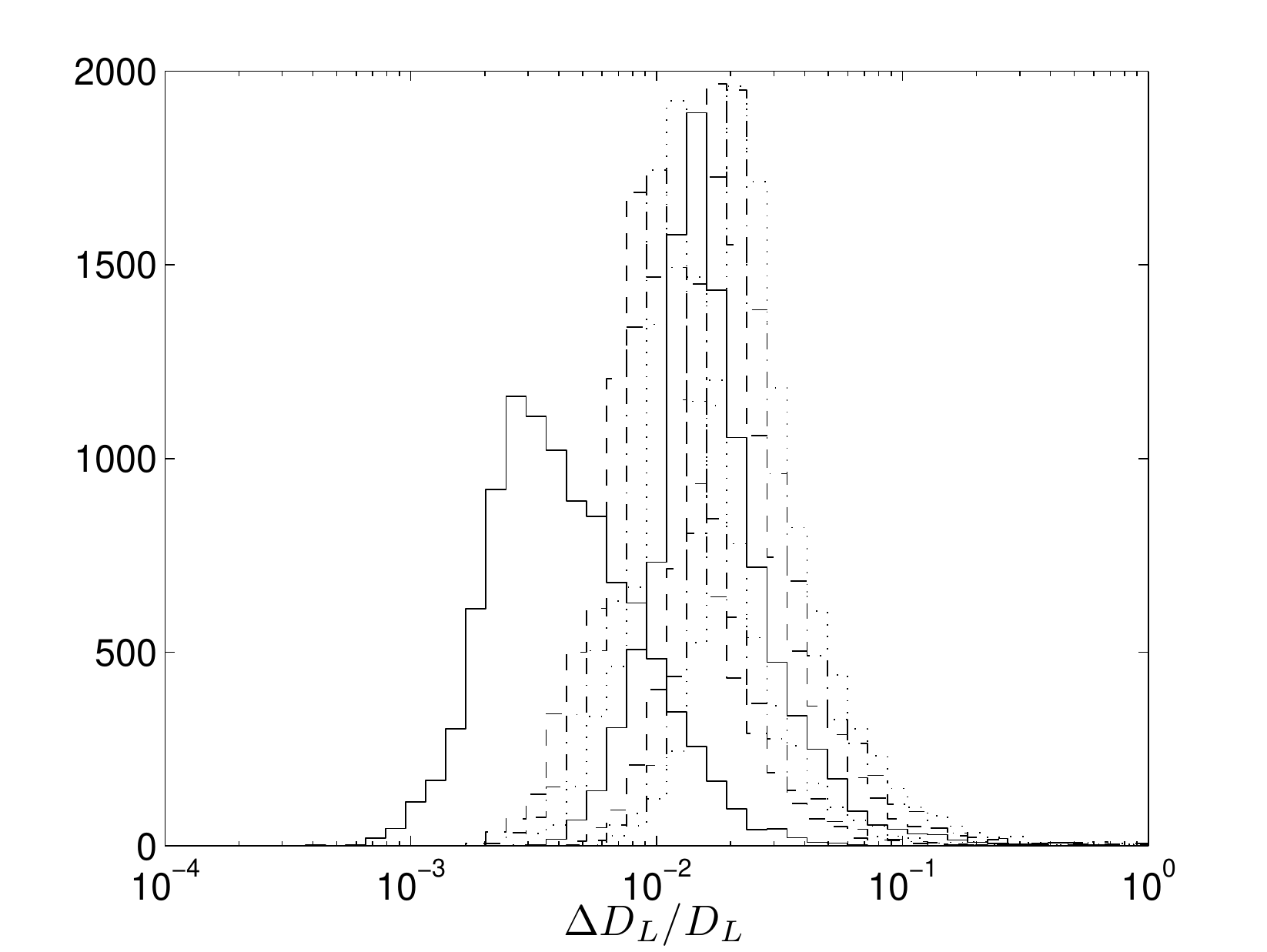}
\caption{Distribution of $\Delta D_L/D_L$ for $10^4$ binaries with $m_1 =
10^6 M_\odot$, $m_2 = 3 \times 10^5 M_\odot$, and $z = 1$ at different
values of $N$ (the number of days before merger).  Reading from left to
right, $N = 0$ ({\it solid line}), 1 ({\it dashed line}), 
2 ({\it dash-dotted line}), 4 ({\it dotted line}), 7 ({\it solid line again}), 
14 ({\it dashed line}), 21 ({\it dash-dotted line}), and
28 ({\it dotted line}).}
\label{fig:Devol}
\end{center}
\end{figure}

Finally, we investigate the time evolution of errors in the luminosity
distance $D_L$.  Figure \ref{fig:Devol} shows the distribution of
$\Delta D_L/D_L$ (determined solely by taking into account GW measurement
effects) evolving in time for a binary with $m_1 = 10^6 M_\sun$, $m_2
= 3 \times 10^5 M_\sun$, and $z = 1$.  We see that in contrast to sky
position, the shape of the distribution does not change very much with
time.  It typically spreads enough to reduce its height, but it maintains
a well-defined peak.  However, the progression of the median is very
similar to the sky position case: It decreases slowly with time until the
last day, when it jumps drastically.  The evolution of median values
of $\Delta D_L/D_L$ follows tracks very similar in shape to those
shown in Figure {\ref{fig:ellipsemedians}}, so we do not show them
explicitly. 

Table {\ref{tab:positionsummary}} summarizes all of our results on the
time evolution of localization.  At low redshift, the ability to
locate an event on the sky is quite good over much of the mass range
even as much as a month in advance of the final merger.  In most
cases, the localization ellipse at $z = 1$ is never larger than about
$10\ \mathrm{deg}^2$ in size, which is comparable to the field of view
of proposed future surveys, such as LSST \citep{t02}.  This
ability degrades fairly rapidly as redshift increases, especially for
larger masses.  At $z = 3$, the ellipse can be $\sim 10\ \mathrm{deg}^2$
a few days in advance of merger for small and intermediate
masses.  At the highest masses, GWs provide very little localization
information.  Going to $z = 5$ makes this even worse; an ellipse of
$\sim 10\ \mathrm{deg}^2$ can be found at most a day prior to merger,
and only for relatively small mass ranges.

In many cases, $\Delta D_L/D_L$ is determined so well by GWs that
gravitational lensing errors are expected to dominate.  As such, the
GW-determined values of $\Delta D_L/D_L$ are essentially irrelevant
for locating these binaries in redshift space; lensing will instead
determine how well redshifts can be measured.  However, in some cases,
the intrinsic distance error exceeds the lensing error, so it is worth
knowing when the ``lensing limit'' can be achieved.  At $z = 1$, the
limit is achieved for most binaries as long as a month before merger.
For $z = 3$, the limit is only achieved a few days to a week
(depending on mass) before merger; for $z = 5$, intrinsic errors
generally exceed the lensing errors even at a day before merger.

\begin{table}[p]
\caption{Summary of time evolution of localization accuracy}
\begin{tabular}{llccccc}
\\
\tableline \tableline
$z$ &
Binary Range &
$N$ (days) &
$2a$ (deg) &
$2b$ (deg) &
$\Delta \Omega_N\ (\mathrm{deg}^2)$&
$\Delta D_L/D_L$ \\

\tableline

$1$ & Low ($M \lesssim 10^6\,M_\odot$) &

$0$ & $(17 - 27)$\tablenotemark{a} & $(6.6 - 13)$\tablenotemark{a} & $0.02 - 0.07$ & $(2.4 - 4)\times 10^{-3}$\\
 &  & $1$ & $(71 - 140)$\tablenotemark{a} & $(26 - 91)$\tablenotemark{a} & $0.3 - 2.8$ & $(8 - 21) \times 10^{-3}$ \\
 &  & $7$ & $(100 - 180)$\tablenotemark{a} & $(54 - 120)$\tablenotemark{a} & $0.9 - 5.1$ & $(1.3 - 3) \times 10^{-2}$ \\
 &  & $28$ & $(140 - 240)$\tablenotemark{a} & $(94 - 180)$\tablenotemark{a} & $2.6 - 9.6$ & $(2 - 4.3) \times 10^{-2}$ \\

\tableline

& Int. ($10^6\,M_\odot \lesssim M \lesssim 4\times 10^6\,M_\odot$) & $0$ & $(31 - 40)$\tablenotemark{a} & $(9.2 - 22)$\tablenotemark{a} & $0.04 - 0.18$ & $(3.8 - 5.6) \times 10^{-3}$\\

& & $1$ & $(56 - 75)$\tablenotemark{a} & $(20 - 48)$\tablenotemark{a} & $0.17 - 0.77$ & $(6.4 - 11) \times 10^{-3}$ \\
 &  & $7$ & $(85 - 110)$\tablenotemark{a} & $(42 - 74)$\tablenotemark{a} & $0.6 - 1.9$ & $(1.1 - 1.7) \times 10^{-2}$ \\
 &  & $28$ & $(130 - 190)$\tablenotemark{a} & $(80 - 130)$\tablenotemark{a} & $2 - 5.4$ & $(1.8 - 3) \times 10^{-2}$ \\

\tableline

& High ($M \gtrsim 6\times 10^6\,M_\odot$) &

$0$ & $0.6 - 1.4$ & $0.2 - 0.64$ & $0.07 - 0.68$ & $(4.6 - 25)\times 10^{-3}$ \\
 &  & $1$ & $1 - 4.7$ & $0.4 - 2$ & $0.2 - 7.1$ & $(0.7 - 16)\times 10^{-2}$ \\
 &  & $7$\tablenotemark{b} & $2.1 - 6.2$ & $0.9 - 3$ & $1 - 14$ & $(1.6 - 8.7)\times 10^{-2}$ \\
 &  & $28$\tablenotemark{b} & $7.1 - 23$ & $3.4 - 8.7$ & $16 - 150$ & $(8.4 - 30)\times 10^{-2}$ \\

\tableline
\tableline

$3$ & Low ($M \lesssim 10^6\,M_\odot$) &

$0$ & $1.3 - 2.4$ & $0.6 - 1.3$ & $0.7 - 2.1$ & $(1.2 - 2)\times 10^{-2}$\\
 &  &  $1$ & $3.9 - 7.2$ & $1.4 - 4.7$ & $3 - 27$ & $(2.7 - 6.3) \times 10^{-2}$ \\
 &  &  $7$ & $5.8 - 9.6$ & $2.9 - 6.5$ & $10 - 51$ & $(4.4 - 9.3) \times 10^{-2}$ \\
 &  & $28$ & $8.5 - 13$ & $5.4 - 9.8$ & $32 - 100$ & $(7.2 - 14) \times 10^{-2}$ \\

\tableline

& Int. ($10^6\,M_\odot \lesssim M \lesssim 4\times 10^6\,M_\odot$) &

$0$ & $2.1 - 2.6$ & $0.7 - 1.4$ & $0.8 - 2.6$ & $(1.5 - 2.4) \times 10^{-2}$\\

& & $1$ & $3.6 - 5$ & $1.3 - 3.1$ & $2.4 - 12$ & $(2.5 - 4.5) \times 10^{-2}$ \\
 &  & $7$ & $6 - 11$ & $2.8 - 6.3$ & $9.8 - 48$ & $(4.5 - 10) \times 10^{-2}$ \\
 &  & $28$ & $9.8 - 91$ & $6.3 - 42$ & $43 - 2900$ & $(0.9 - 14) \times 10^{-1}$ \\

\tableline

& High ($M \gtrsim 6\times 10^6\,M_\odot$) &

$0$ & $3.4 - 33$ & $1.3 - 9.7$ & $2.7 - 260$ & $(2.9 - 240)\times 10^{-2}$ \\
 &  & $1$\tablenotemark{b} & $5.7 - 17$ & $2.4 - 7.3$ & $8.5 - 93$ & $(5.2 - 53)\times 10^{-2}$ \\
 &  & $7$\tablenotemark{b} & $25 - 75$ & $13 - 27$ & $220 - 1500$ & $(5.8 - 19)\times 10^{-1}$ \\
 &  & $28$\tablenotemark{c} & $\cdots$ & $\cdots$ & $\cdots$ & $\cdots$ \\

\tableline
\tableline

$5$ & Low ($M \lesssim 10^6\,M_\odot$) &

$0$ & $2.8 - 4.9$ & $1.1 - 2.7$ & $2.3 - 9.3$ & $(2.6 - 4.1)\times 10^{-2}$\\
 &  & $1$ & $6.8 - 12$ & $2.4 - 7.6$ & $9 - 71$ & $(4.6 - 10) \times 10^{-2}$ \\
 &  & $7$ & $10 - 16$ & $5 - 11$ & $32 - 140$ & $(7.8 - 15) \times 10^{-2}$ \\
 &  & $28$ & $16 - 22$ & $9.6 - 16$ & $100 - 290$ & $(1.3 - 2.3) \times 10^{-1}$ \\

\tableline

& Int. ($10^6\,M_\odot \lesssim M \lesssim 4\times 10^6\,M_\odot$) &

$0$ & $3.9 - 5.2$ & $1.4 - 2.7$ & $3.3 - 10$ & $(3.2 - 5)\times 10^{-2}$ \\

& & $1$ & $7.2 - 11$ & $2.5 - 6.2$ & $9.6 - 48$ & $(5.2 - 10)\times10^{-2}$ \\
 &  & $7$ & $12 - 41$ & $5.9 - 21$ & $45 - 610$ & $(1.1 - 5.2)\times10^{-1}$ \\
 &  & $28$\tablenotemark{b} & $21 - 170$ & $15 - 65$ & $250 - 8000$ & $(2 - 21)\times10^{-1}$ \\

\tableline

& High ($M \gtrsim 6\times 10^6\,M_\odot$) &

$0$\tablenotemark{b} & $9 - 29$ & $4.3 - 10$ & $30 - 230$ & $(1.2 - 12)\times10^{-1}$ \\
 &  & $1$\tablenotemark{b} & $19 - 27$ & $9.9 - 12$ & $130 - 250$ & $(4.4 - 7.3)\times10^{-1}$ \\
 &  & $7$\tablenotemark{c} & $\cdots$ & $\cdots$ & $\cdots$ & $\cdots$ \\
\tableline
\end{tabular}

\tablecomments{Typical ranges of sky position and distance measurement accuracy as a function
of time until merger for low-, intermediate-, and high-mass binaries.  Angles are in degrees unless otherwise noted;
solid angles are in square degrees.}

\tablenotetext{a}{These angles are in arcminutes; all others
are in degrees.}

\tablenotetext{b}{Some very massive systems are excluded from these
data.  In those cases, the position and distance are very poorly constrained this far in advance of merger.  In some cases, the binary is even out of band.}

\tablenotetext{c}{All of the binaries of this mass and redshift are either very poorly measured or completely out of band this far in advance of merger.}

\label{tab:positionsummary}
\end{table}

\subsection{Results II: Angular dependence of localization accuracy}
\label{sec:angdependence}

We now examine one final interesting property of the errors: their
dependence on the sky position of the source.  As we design future
surveys to find counterparts to MBHB coalescences, it will be
important to understand if there is a bias for good (or bad)
localization in certain regions of the sky.  It is also useful to know
in advance whether the ``best'' regions are likely to be blocked by
foreground features such as the Galactic center.

Before discussing this dependence in detail, it is worth reviewing
some details of how our Monte Carlo distributions are constructed.  As
described in \S\ {\ref{sec:review}}, in most of our analysis we
randomly distribute the sky position of our binaries, drawing from a
uniform distribution in $\bar{\mu}_N = \cos\bar{\theta}_N$ and
$\bar{\phi}_N$, where $\bar{\theta}_N$ and $\bar{\phi}_N$ are the
polar and azimuthal angles of a binary in solar system barycenter
coordinates.  We also randomly choose our binaries' final merger time.
In this section, rather than distributing the sky position, we examine
parameter accuracies for particular given positions; all other Monte
Carlo parameters are distributed as usual.  Because we continue to
randomly distribute the final merger time, the {\it relative} azimuth
between a binary's sky position and {\it LISA}'s orbital position at
merger, $\delta\phi = \bar{\phi}_N - \phi_{\it LISA}(t_c)$, remains
randomly distributed.  As such, we expect our analysis to effectively
average over $\bar{\phi}_N$, washing out any strong dependence on this
angle in our analysis.

\begin{figure}[t]
\includegraphics[scale=0.32]{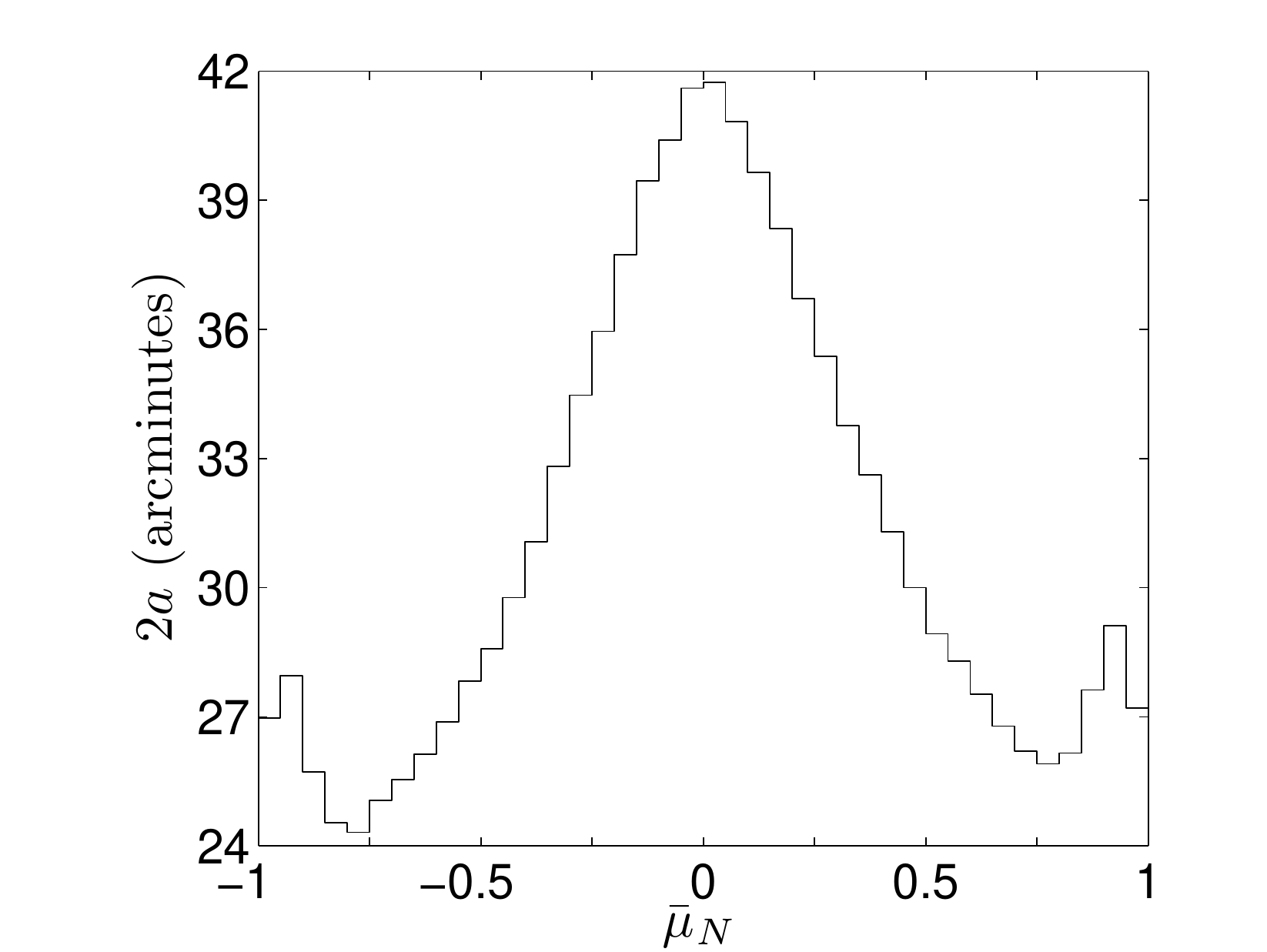}
\includegraphics[scale=0.32]{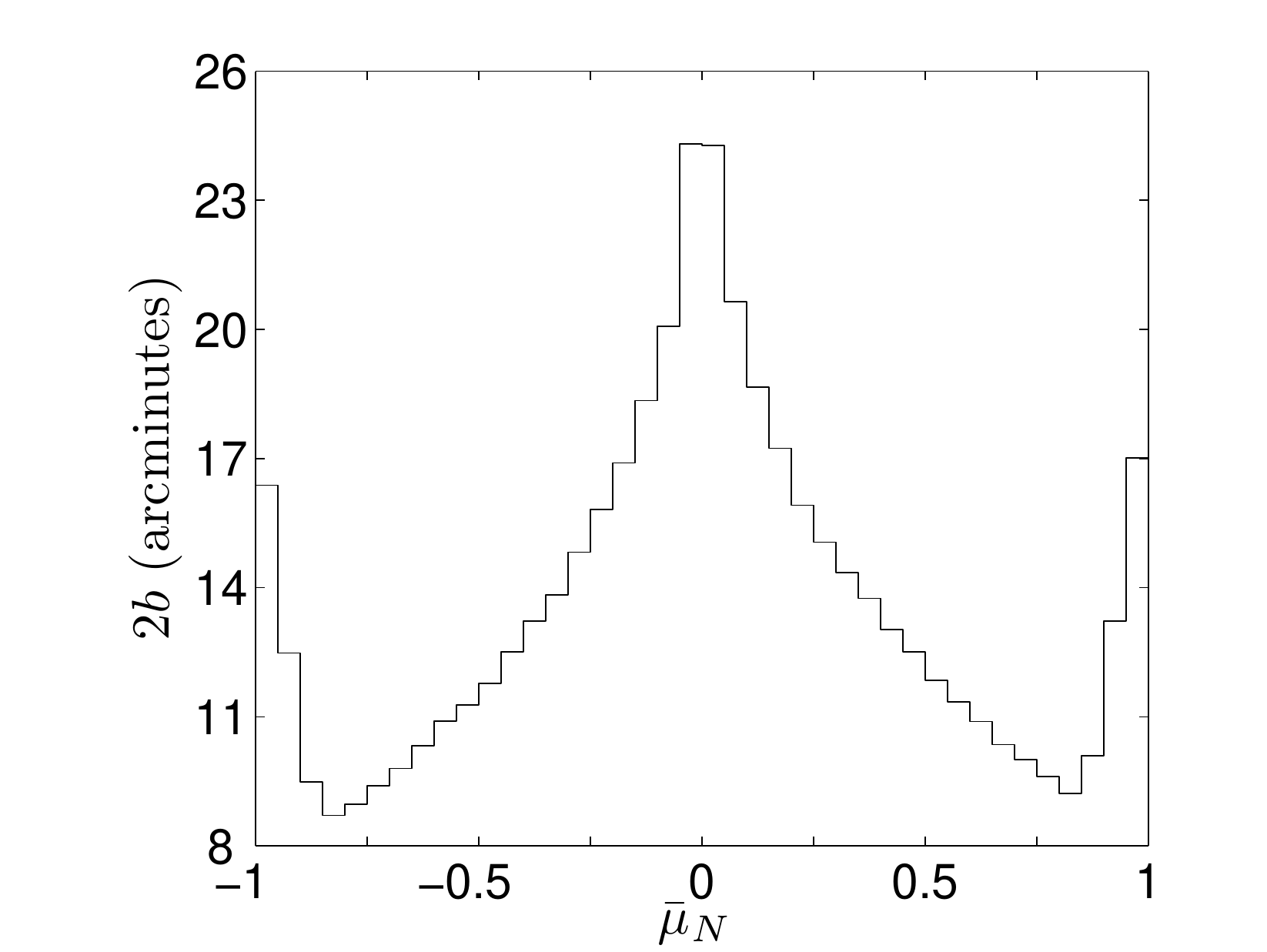}
\includegraphics[scale=0.32]{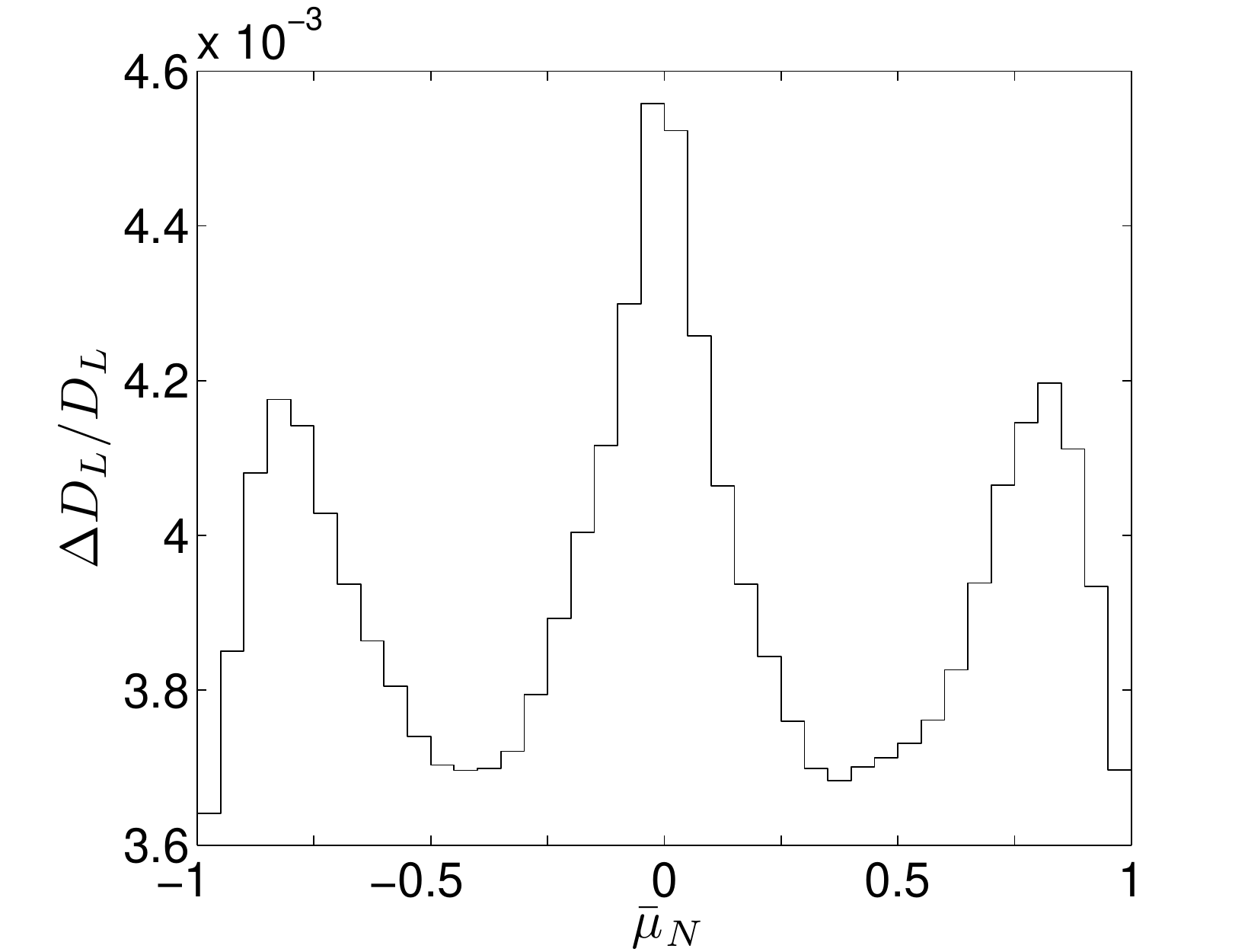}
\caption{Dependence of the localization errors on $\bar{\mu}_N$.
The major axis $2a$ of the sky position error ellipse is on the left, the 
minor axis $2b$ in the center, and the 
luminosity distance errors $\Delta D_L/D_L$ on the right.  Each datum
represents the median of $10^4$ binaries with $m_1 = 10^6 M_\sun$,
$m_2 = 3 \times 10^5 M_\sun$, and $z = 1$; all other parameters are
selected randomly (except for $\bar{\mu}_N$, whose range is limited to
the bin width).}
\label{fig:mudependence}
\end{figure}

We begin by examining the dependence of errors on $\bar{\mu}_N$.  We
evenly divide the range $-1 \le \bar{\mu}_N \le 1$ into 40 bins and
run a Monte Carlo simulation with $10^4$ points in each.  That is, we
pick $\bar{\mu}_N$ only from the bin range, but we pick all other
random parameters in the usual manner.  The results for a
representative binary ($m_1 = 10^6 M_\odot, m_2 = 3 \times 10^5
M_\odot$, and $z = 1$) are shown in Figure {\ref{fig:mudependence}}.

Note that all of the error distributions are symmetrically peaked
around the plane of {\em LISA}'s orbit ($\bar\mu_N = 0$).  Any slight
asymmetry is due only to statistical effects.  This is reassuring;
{\em LISA} should not favor one hemisphere over the other.  There is
also additional structure that is parameter dependent.  At its peak,
the major axis $2a$ is almost $35 \%$ greater than the
position-averaged median value\footnote{Note that the
position-averaged medians quoted here are slightly different from
those quoted in Table 1, since this sample has 40 times more points.}
of $31.1^\prime$.  It then decreases with $|\bar{\mu}_N|$ and
reaches a minimum of about $25^\prime$ for $0.75 < |\bar{\mu}_N| <
0.8$.  Finally, there are subpeaks near the ecliptic poles, although
they still lie below the position-averaged median.  The dependence of
the minor axis $2b$ on angle is even more dramatic.  At its peak, $2b$
differs from the position-averaged median of $13.0^\prime$ by over
$85\%$.  Just like the major axis, it drops to a minimum, but it does
so more rapidly.  The minimum also occurs at a slightly larger value of $|\bar{\mu}_N|$
than for $2a$.  The minor axis also shows fairly strong subpeaks near
the ecliptic poles, with values higher than the position-averaged
median.

The luminosity distance errors behave slightly differently.  First of
all, the variation with $\bar{\mu}_N$ is weaker than for the sky
position: At the central peak, $\Delta D_L/D_L$ is only $\sim 15 \%$
greater than its position-averaged median value ($0.00392$).  In
addition, while the distribution again peaks near the poles, it does
so at a smaller value of $|\bar{\mu}_N|$.  (In fact, the peaks occur
very close to where sky position errors are {\it minimized}.)

By binning the data sets developed for Paper I, we were able to
confirm this behavior over a wide range of masses and redshifts,
albeit with poorer statistics: $10^4$ binaries in total, rather
than per bin.  Thanks to the poorer statistics, we cannot
resolve the small polar subpeaks in the distribution of $2a$.  In addition, in
some cases it appears that the side peaks can be larger than the
central peak in the distribution of $\Delta D_L/D_L$.  Aside from
these minor variations, the shapes and relative amplitudes seen in
these distributions hold robustly over all masses and redshifts we
considered.


\begin{figure}[!t]
\includegraphics[scale=0.32]{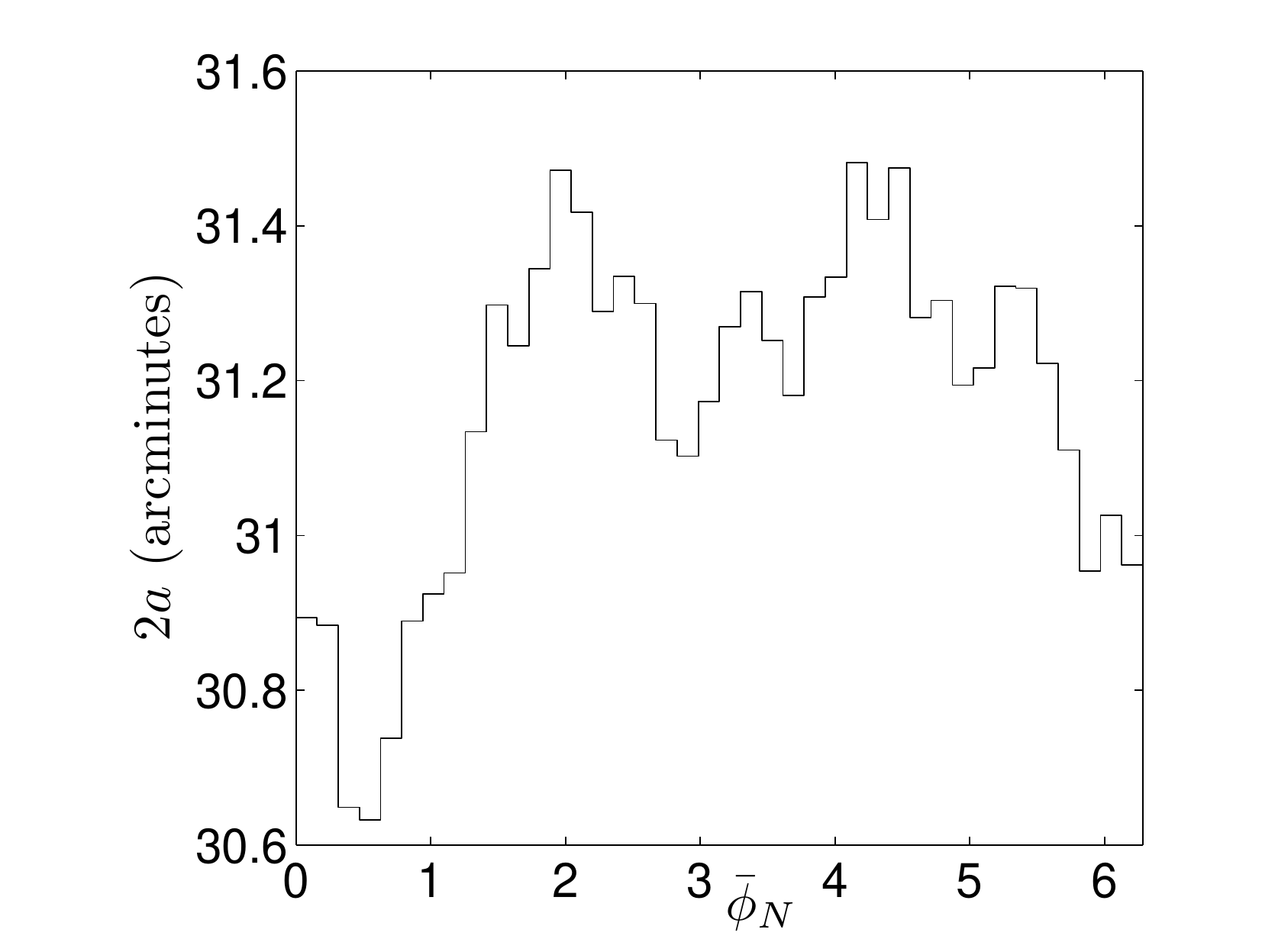}
\includegraphics[scale=0.32]{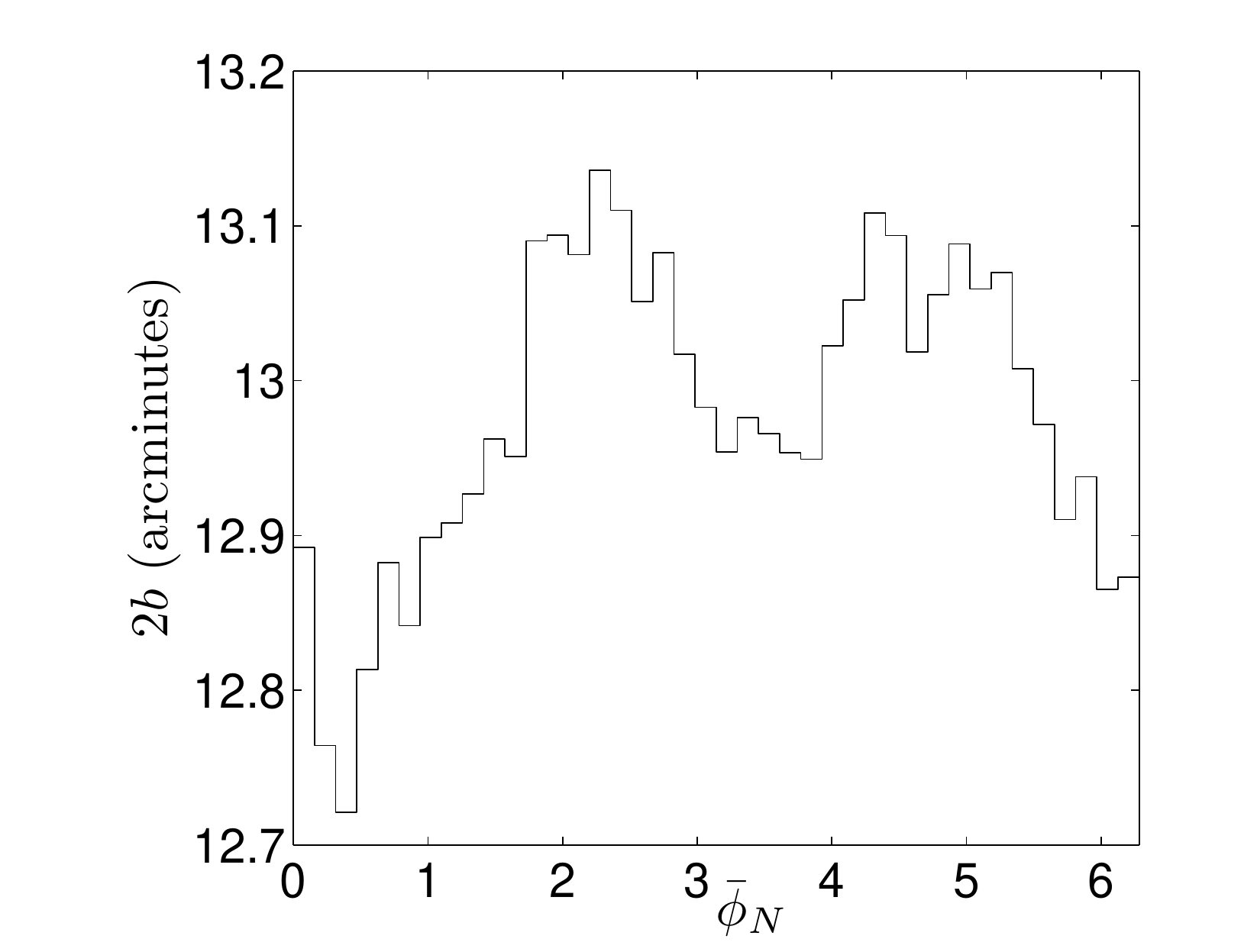}
\includegraphics[scale=0.32]{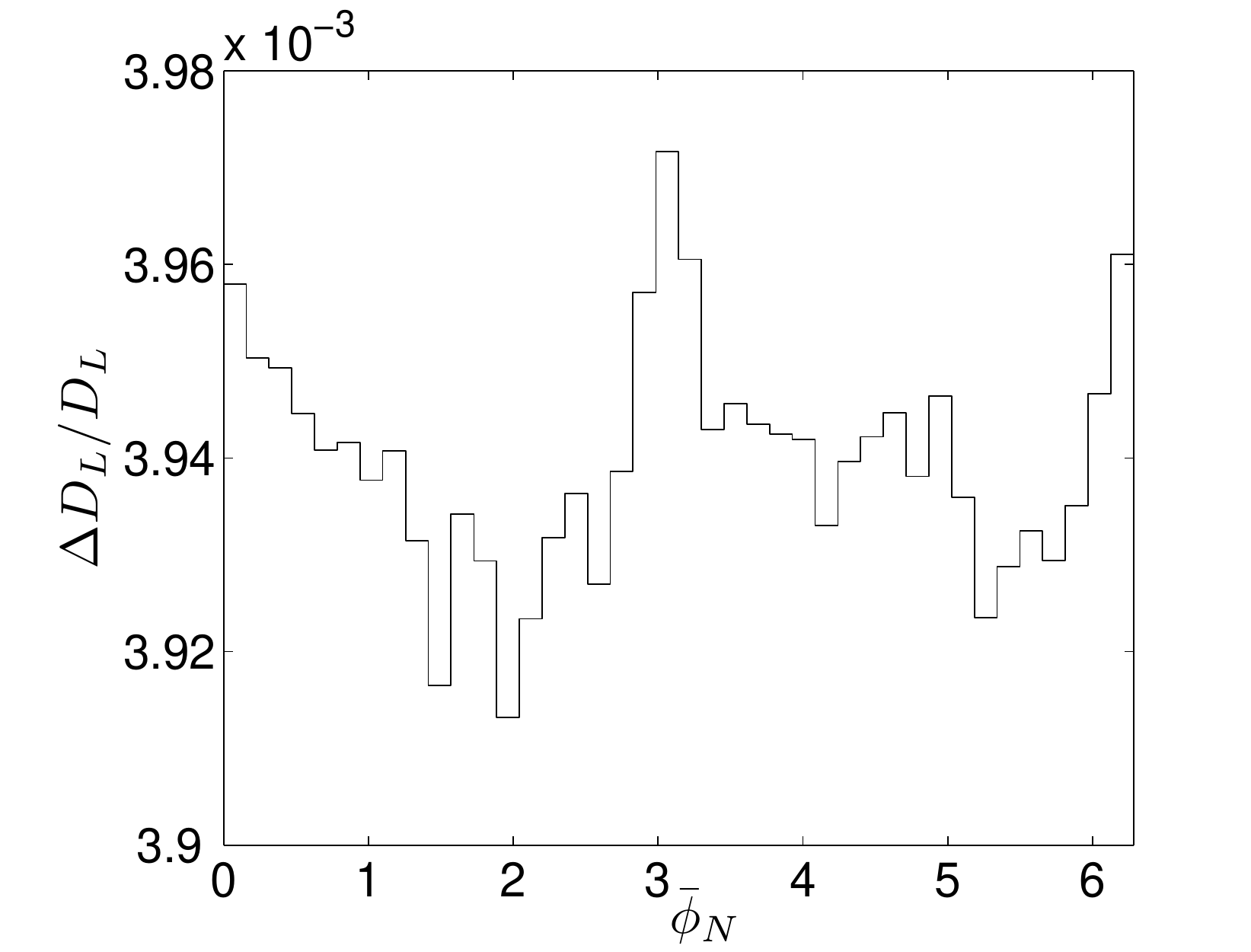}
\caption{Same as Fig. \ref{fig:mudependence}, but for dependence on $\bar{\phi}_N$ and with $10^5$ binaries in each bin rather than $10^4$.}
\label{fig:phidependence}
\end{figure}

We next investigate the dependence of the errors on the azimuthal
angle $\bar{\phi}_N$.  To improve the statistics, we now calculate
$10^5$ binaries in each bin; the results are shown in Figure
\ref{fig:phidependence}.  The errors have a very weak (although
nonzero) dependence on $\bar{\phi}_N$: The maximum deviation from the
overall median is only $\sim 1\% - 2\%$.  This is to be expected; as
discussed at the beginning of this section, the randomness of our
binaries' merger times effectively averages over azimuth.  If we did
not average over azimuth in this way, we would expect a moderately
strong $\bar{\phi}_N$-dependence due to the functional form of {\it
LISA}'s response to GWs.  Even after averaging this dependence away,
we might expect some residual $\bar{\phi}_N$ structure due to the
``rolling'' motion of the {\it LISA} constellation.  The phase
associated with {\it LISA}'s roll angle puts an additional oscillation
on the measured waves (see the $\alpha$-dependence in eqs.\ [47] and
[48] of K07, where $\alpha$ encodes the roll angle), and we do not
average over this angle.  Indeed, on close inspection, we can make
out roughly two peaks in each plot in Figure \ref{fig:phidependence}
(although the statistics are still too poor to resolve them clearly),
consistent with the $\cos2\alpha$ and $\sin2\alpha$ behavior shown in
K07.  For comparison, we also examined the $\bar{\phi}_N$-dependence for
this mass and redshift with precession ``turned off.''  The
oscillatory behavior appears very clearly in this case.  Because of
the weakness of the $\bar{\phi}_N$ behavior, we were unable to easily
check it for other masses and redshifts.

\begin{figure}[!p]
\begin{center}
\includegraphics[scale=0.48]{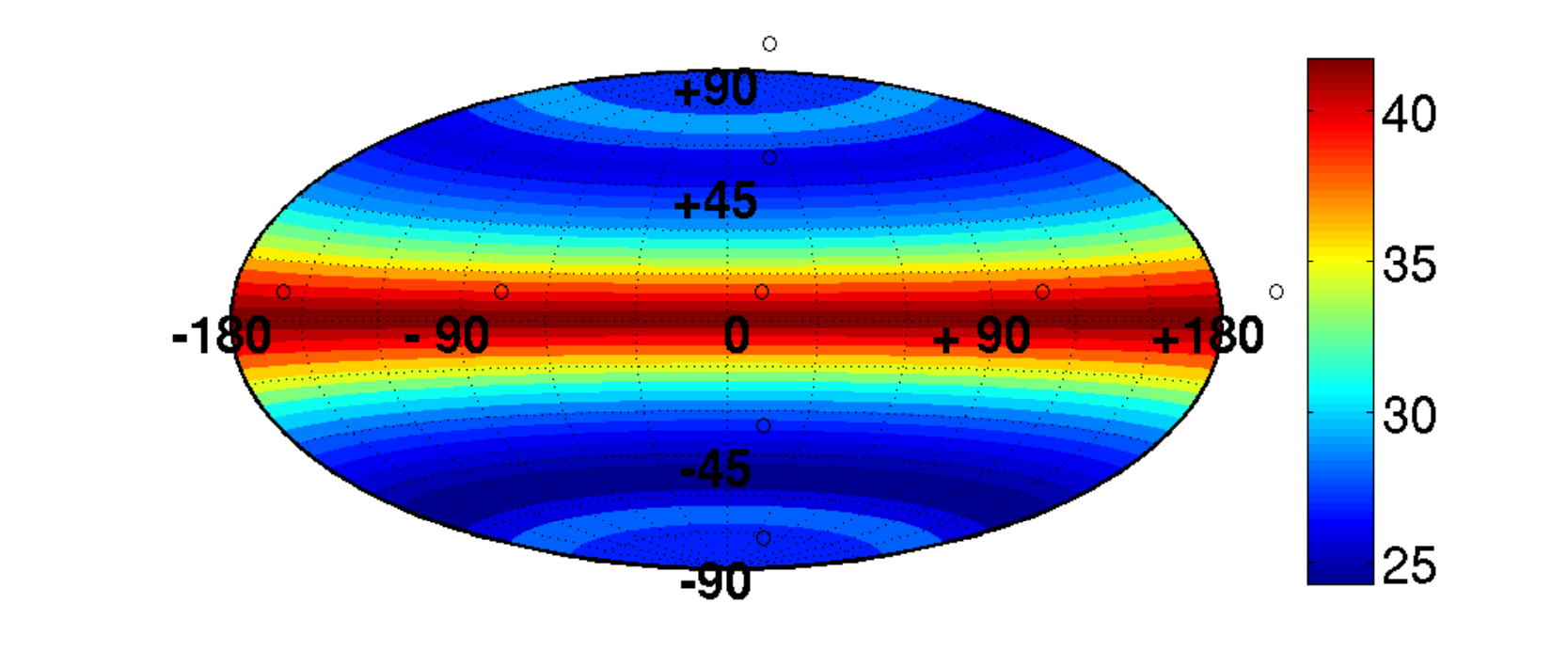}
\includegraphics[scale=0.48]{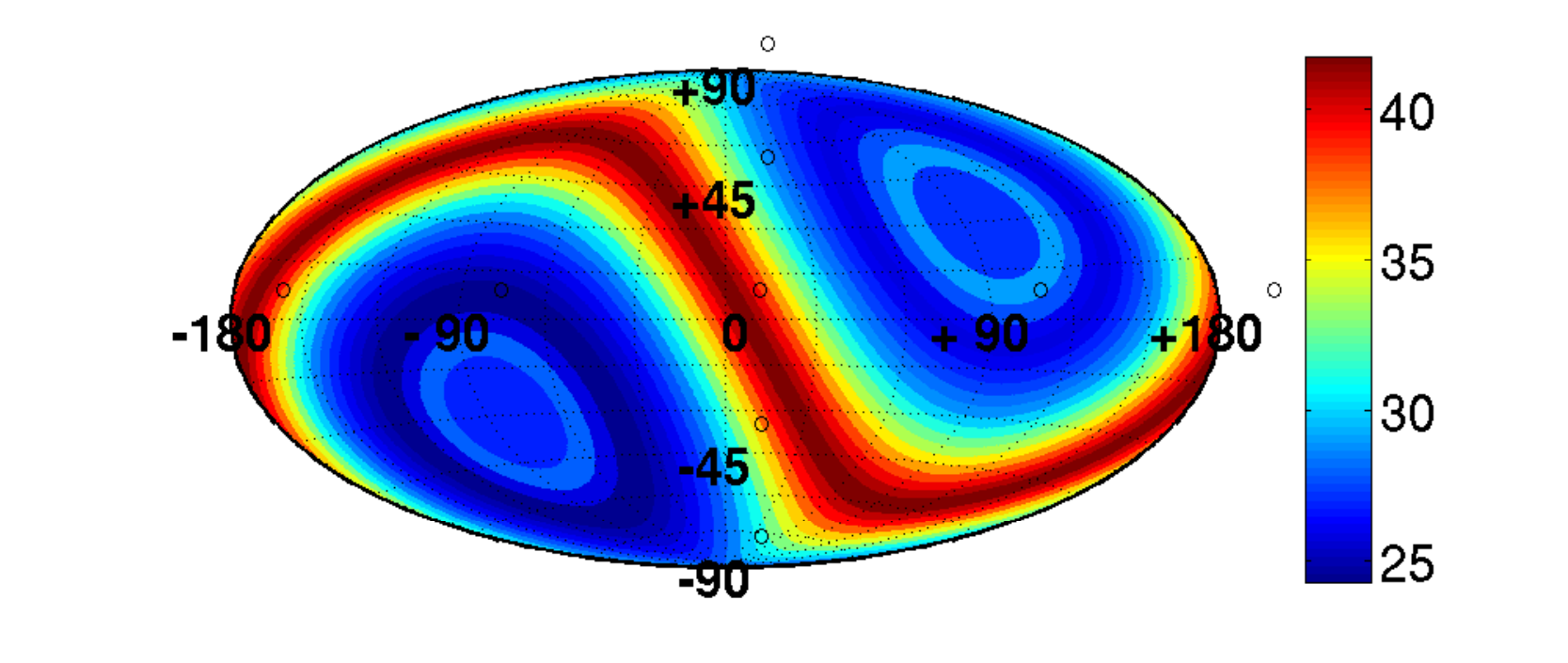}
\includegraphics[scale=0.48]{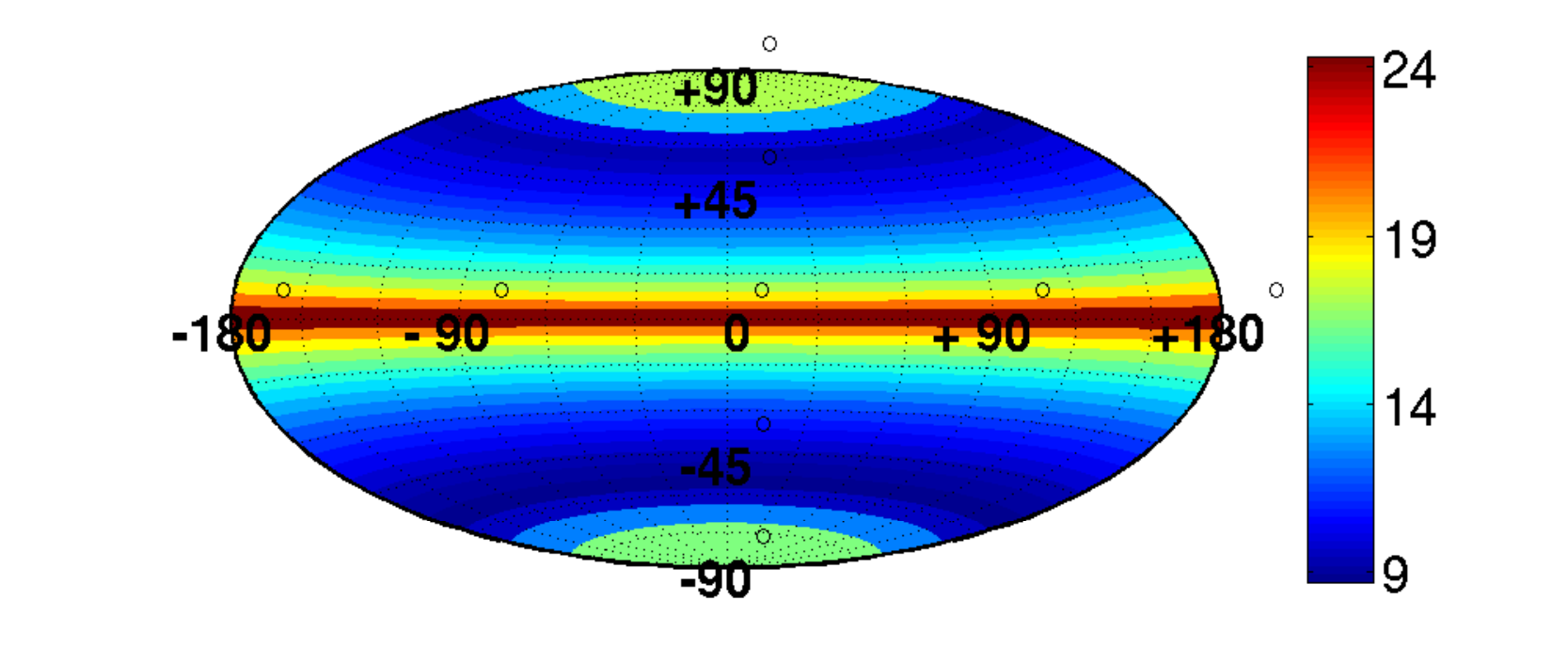}
\includegraphics[scale=0.48]{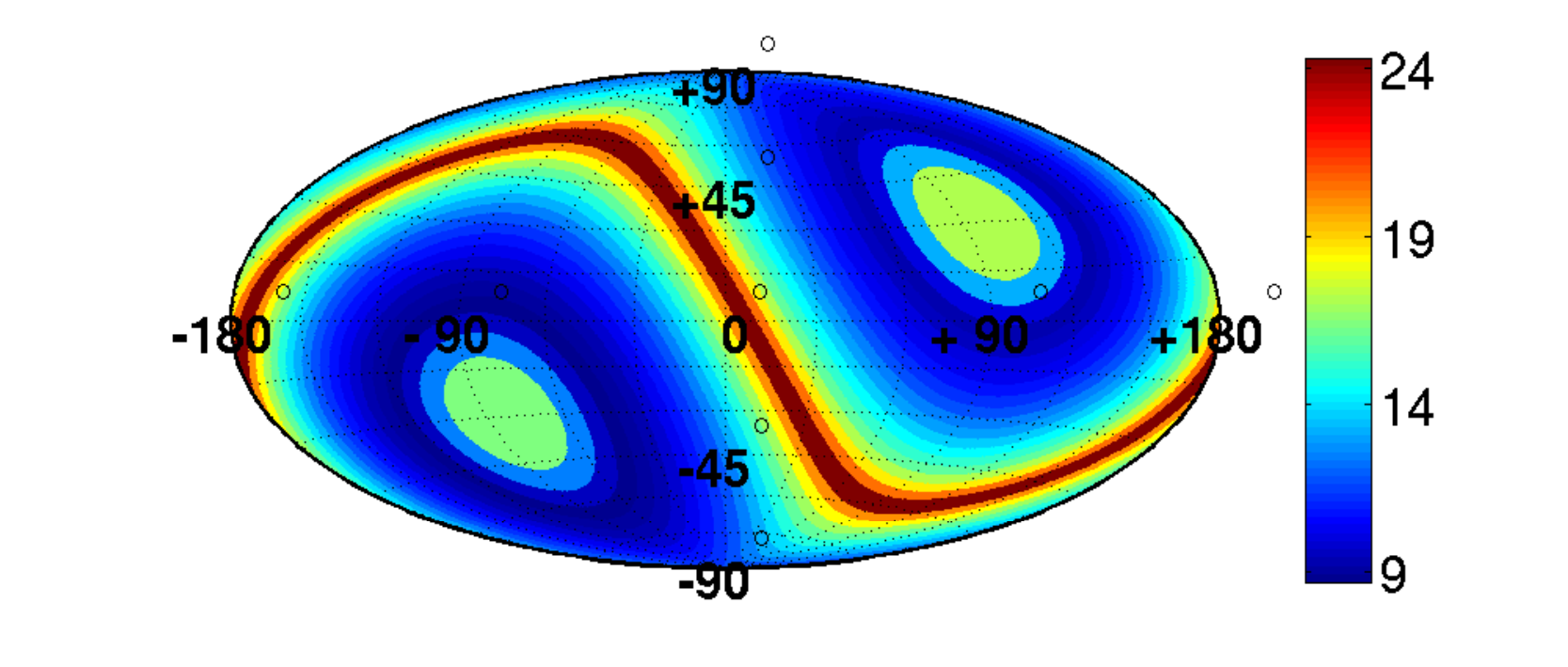}
\includegraphics[scale=0.48]{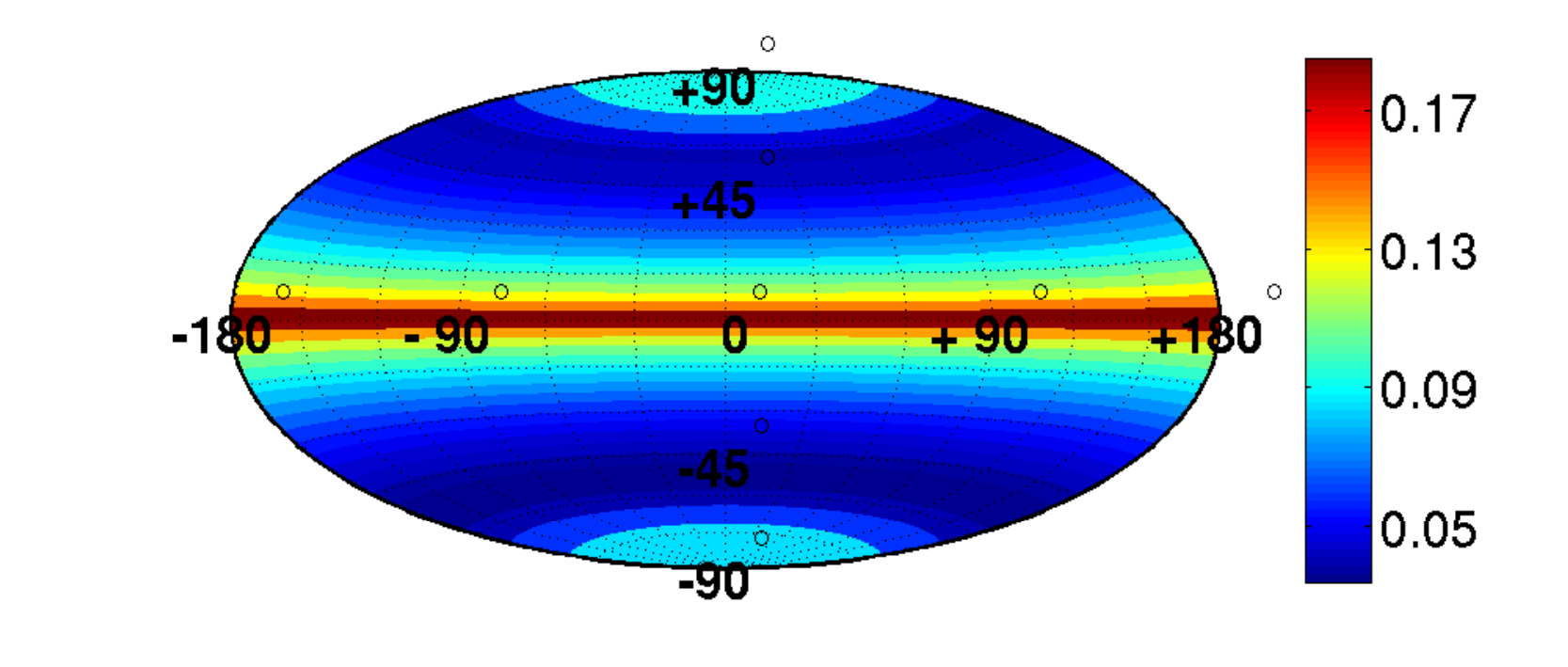}
\includegraphics[scale=0.48]{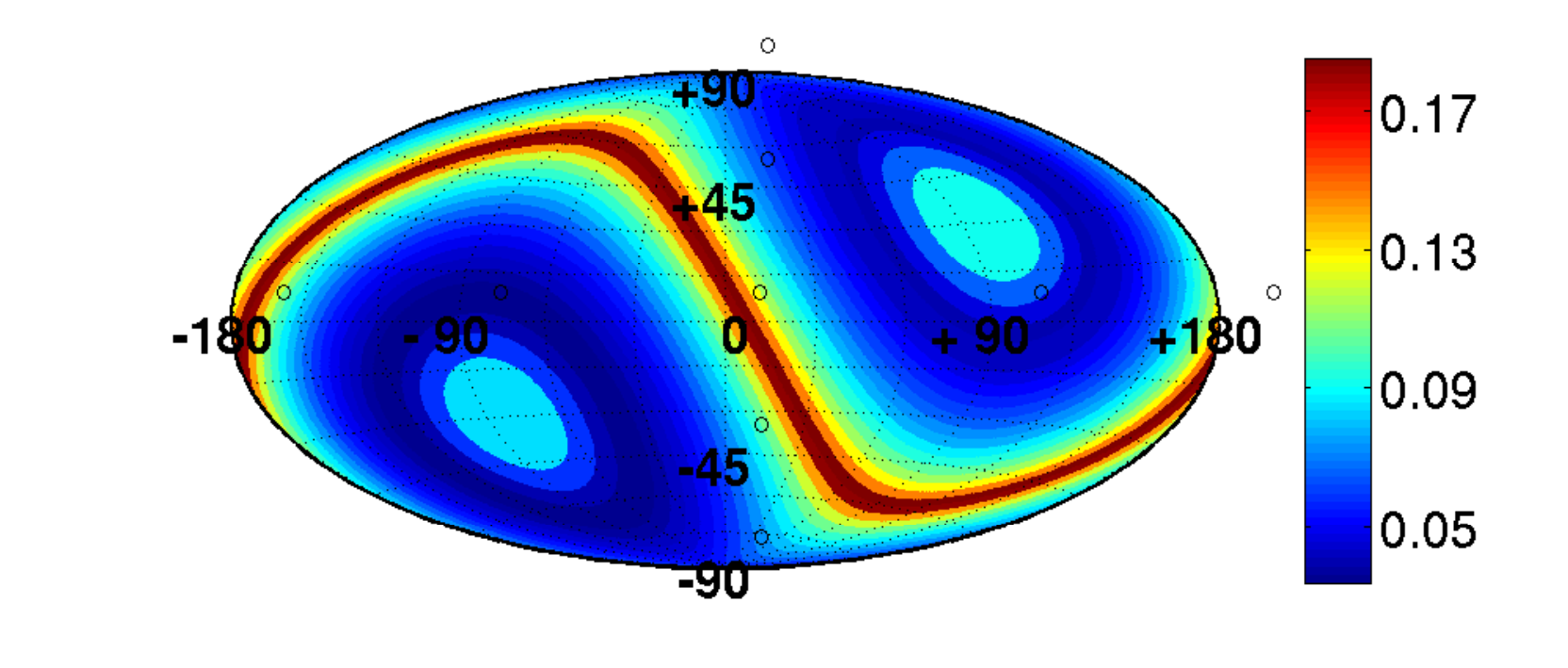}
\includegraphics[scale=0.48]{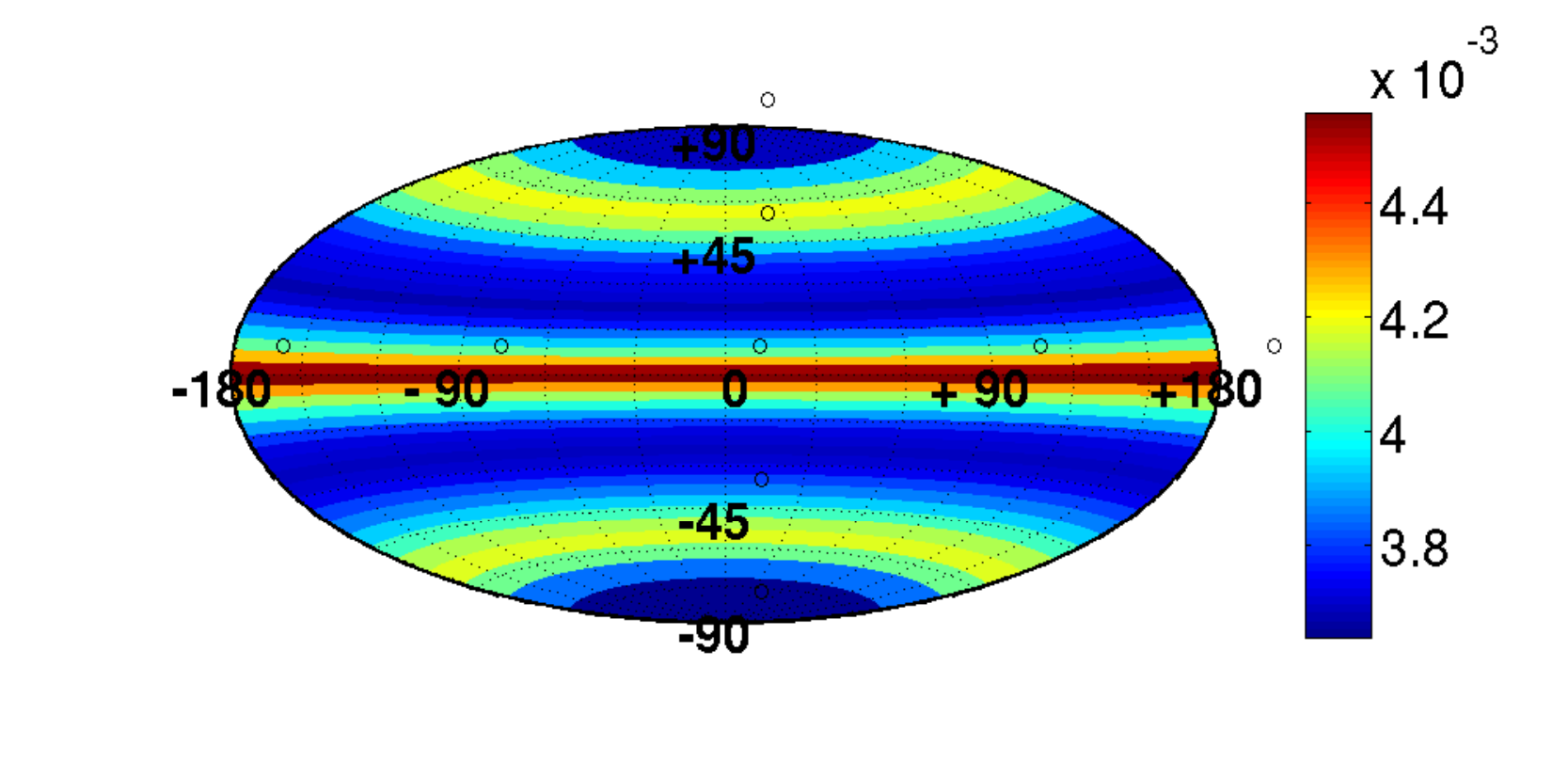}
\includegraphics[scale=0.48]{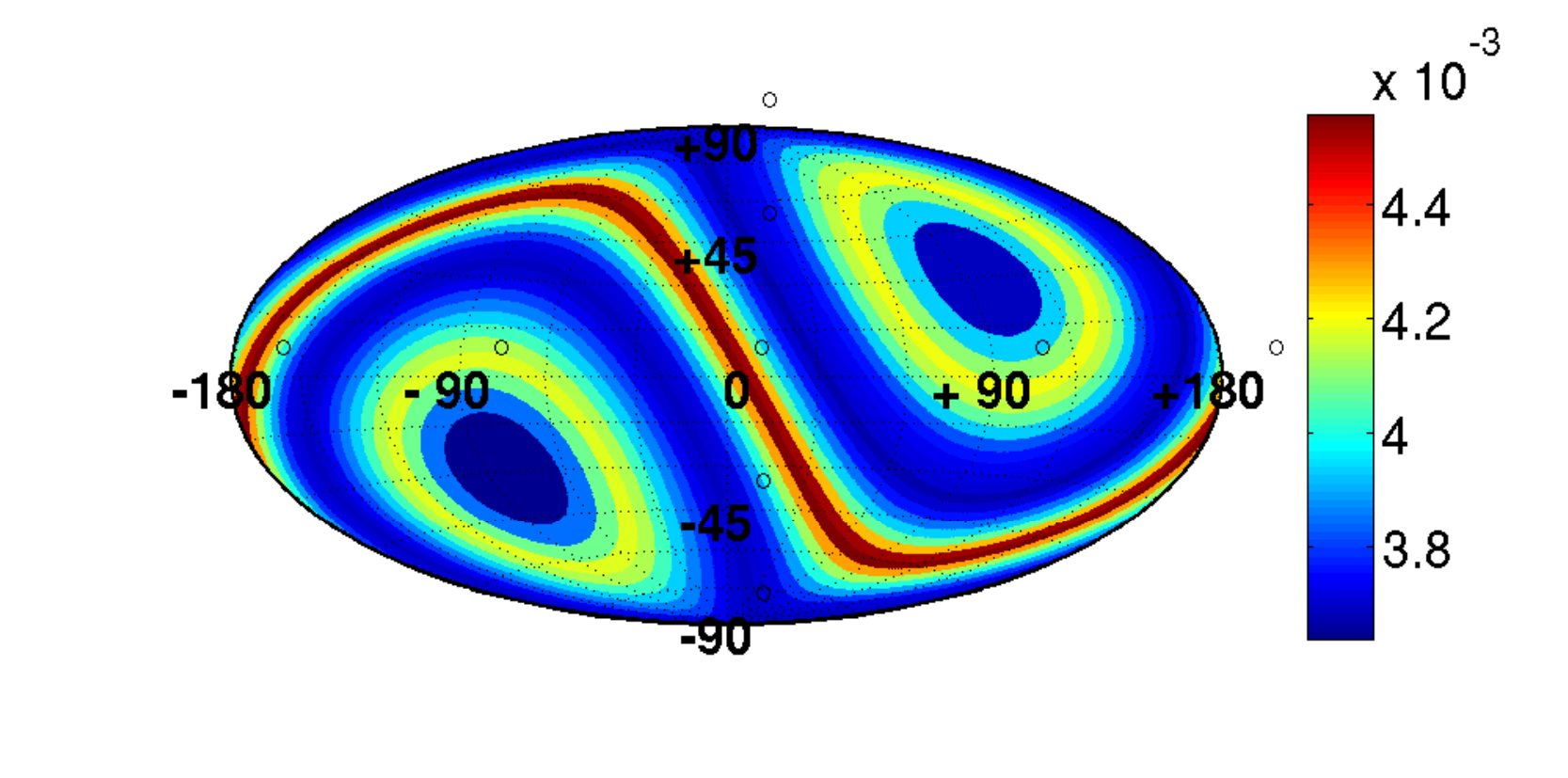}
\caption{Sky maps of major axis $2a$ ({\it top row}, in arcminutes), minor axis
$2b$ ({\it second row}, in arcminutes), localization ellipse area $\Delta
\Omega_N$ ({\it third row}, in square degrees) and $\Delta D_L/D_L$ 
({\it bottom row}) for {\em LISA} observations of MBHBs
in different parts of the sky.  Data in the left column are 
presented in ecliptic coordinates;
data on the right are in Galactic coordinates, with the Galactic
center at the middle.  Note that the level of $\bar\phi_N$ variation
is so small that it would not show up in these figures; accordingly,
they are essentially just remappings of Fig.\
{\ref{fig:mudependence}}.}
\label{fig:skymaps}
\end{center}
\end{figure}

Finally, to give an overall sense as to how localization varies on the
sky, we present in Figure {\ref{fig:skymaps}} sky maps of the median
major axis $2a$, minor axis $2b$, localization ellipse area $\Delta
\Omega_N$, and distance accuracy $\Delta D_L/D_L$.  We show these data
both in ecliptic and Galactic coordinates.  Note that most of the
region of small error lies outside of the Galactic plane.  This
potentially bodes well for searches for MBHB electromagnetic
counterparts --- the regions where instruments like {\it LISA} ``see''
most sharply are less likely to be hidden by foreground features.
Certain portions of the sky that {\it LISA} sees well will be easier
to search telescopically than others.  It will be an important task
for future surveys over all electromagnetic bands to identify regions
that are particularly amenable to finding counterparts to MBHB events.

\section{Summary and conclusions}
\label{sec:disc}

As discussed at length in Paper I, accounting for the general
relativistic precession of the angular momentum vectors in an MBHB system has a
dramatic impact on what we can learn by observing the system's
gravitational waves.  Spin-induced precession breaks degeneracies
among different parameters, making it possible to measure them more
accurately than they could be determined if precession were not
present.  This has a particularly important impact on our ability to
locate such a binary on the sky and to determine its luminosity
distance --- the degeneracy between sky angles, distance, and
orientation angles is severe in the absence of precession.

Our analysis shows that the improvement that precession imparts to
measurement accumulates fairly slowly.  In using one code which includes
the impact of spin precession and a second which neglects this effect,
we find little difference in the accuracy with which GWs determine
sky position and distance for times more than a few days in advance of the
final merger.  The difference between the two codes grows quite
rapidly in these final days.  In the last day alone, the localization 
ellipse area decreases by a factor of $\sim 3-10$ (up to $\sim 60$ in a 
few low-mass systems) when precession effects are included.  
Distance determination is likewise improved by
factors of $\sim 1.5-7$ in that final day.  

Not all of the precession effects occur in the final days.  We saw in
Figure \ref{fig:axes_evol} that the long tail of small minor axes can be
seen, to some degree, throughout the inspiral.  We could get lucky and
find a binary with a very small value of $2b$ weeks before merger.
But the improvement in the median that we found in Paper I appears to
take effect only in the final days of inspiral.  Therefore, while
precession may in fact help improve the {\it final} localization of a
coalescing binary by a factor of $\sim 2-10$ in each direction, it
will not be much help in {\it advanced} localization of a typical
binary.

Nevertheless, the pixel sizes that we find are small enough that
future surveys should not have too much trouble searching the region
identified by GWs, at least over certain ranges of mass and redshift.
At $z = 1$, the GW localization ellipse is $\sim 10\ \mathrm{deg}^2$ or
smaller for most binaries as early as a month in advance of merger.
(At high masses, the ellipse can be substantially larger than this a
month before merger, but it shrinks rapidly, reaching a comparable
size $1-2$ weeks before merger.)  This bodes well for future surveys
with large fields of view that are likely to search the GW pixel for
counterparts.  In addition, GWs determine the source luminosity
distance so well that the distance errors we find are essentially
irrelevant --- gravitational lensing will dominate the distance error
budget for all but the highest masses.

As redshift increases, the GW pixel rapidly degrades, particularly for the
largest masses.  Let us adopt $10\ \mathrm{deg}^2$ (the approximate LSST
field of view) as a benchmark localization for which counterpart
searches may be contemplated.  At $z = 3$, this benchmark is reached
at merger for almost the entire range of masses we considered.  As
little as a day in advance of merger, however, some of the least
massive and most massive systems are out of this regime.  One week
prior to merger, the most massive systems are barely located at all
(ellipses hundreds of square degrees or larger).  The intermediate
masses do best, but even in their cases the positions are determined
with $\sim 10\ \mathrm{deg}^2$ accuracy no earlier than a few days in
advance of merger.  The resolution degrades further at higher
redshift.  At $z = 5$, systems with $M \gtrsim 6 \times 10^6\,M_\odot$
are not located more accurately than $\sim 30\ \mathrm{deg}^2$ even at
merger.  Smaller systems are located within $\sim 10\ \mathrm{deg}^2$
at merger, but very few are at this accuracy even one day in advance
of merger.  The luminosity distance errors also increase, so much that
they exceed lensing errors a few days to a week before merger at $z =
3$, and only a day before merger at $z = 5$.  This degradation hurts
the ability to search for counterparts by redshift and subsequently
use them as standard candles.

Our main conclusion is that future surveys are likely to have good
advanced knowledge (a few days to one month) of the location of MBHB
coalescences at low redshift ($z \sim 1 - 3$), but only a day's notice
at most at higher redshift ($z \sim 5$).  This conclusion may be
excessively pessimistic.  As mentioned earlier, recent work examining
the importance of subleading harmonics of MBHB GWs is finding that
including harmonics beyond the leading quadrupole has an important
effect on the final accuracy of position determination
{\citep{arunetal,ts08}}.  For most masses, these analyses show a
factor of a few improvement in position, comparable to the improvement
that we find when spin precession is added to the waveform model.  For
high-mass systems, the higher harmonics increase the (previously
small) overlap with the {\it LISA} band; consequently, the improvement
can be much larger, up to 2 or 3 orders of magnitude in area.
Since these two improvements arise from very different physical
effects, it is likely that their separate improvements can be combined
for an overall improvement significantly better than each effect on
its own.  We plan to test this in future work (which is just now
getting underway).

Finally, we have also studied the sky position dependence of {\it
LISA}'s ability to localize sources.  We have found that the regions
of best localization lie fairly far out of the Galactic plane.
However, as emphasized by N. Cornish (2007, private communication), a proper
anisotropic confusion background might impact this dependence.  In our
calculations, we have assumed an isotropic background, neglecting the
likely spatial distribution of Galactic binaries.  Properly accounting
for this background is likely to strengthen our conclusion that LISA's
ability to ``see'' is best for MBHB sources out of the Galactic plane.

\acknowledgments

We are very grateful to Bence Kocsis for detailed discussions on this
work.  We have also benefitted from discussions with Emanuele Berti,
Neil Cornish, Zoltan Haiman, Bala Iyer, Kristen Menou, B.\
Sathyaprakash, Rob Simcoe, Alicia Sintes, Michele Vallisneri, and
Alberto Vecchio.  This work was supported by NASA grants NAGW-12906
and NNG05G105G, as well as NSF grant PHY04-49884.  S.A.H. gratefully
acknowledges the support of the MIT Class of 1956 Career Development
fund, as well as the hospitality of the Harvard-Smithsonian Center for
Astrophysics' Institute for Theory and Computation, where a portion of
an early draft of this manuscript was written.

\end{document}